\documentclass[a4paper,10pt,twocolumn]{IEEEtran}
\usepackage{color,amsfonts,lettrine,amsmath,amssymb,colortbl,booktabs,multirow,psfrag,bm,amsbsy,cite,multicol,lipsum,etoolbox}

\usepackage{relsize}

\newtoggle{2Collumn}
\toggletrue{2Collumn}

\makeatletter
\def\vect{\mathop{\operator@font vec}\nolimits}
\makeatother
\usepackage{algorithm,algcompatible}
\algnewcommand\INPUT{\item[\textbf{Input:}]}%
\algnewcommand\OUTPUT{\item[\textbf{Output:}]}%

\usepackage{enumerate}
\usepackage{subfigure}
\ifCLASSINFOpdf
   \usepackage[pdftex]{graphicx}

  \usepackage{epstopdf}
  \DeclareGraphicsExtensions{.pdf,.jpeg,.png}\else  
	\DeclareMathOperator{\Tr}{Tr}
\fi

\DeclareMathOperator{\Tr}{Tr}

\begin{document}

%
\title{ Reciprocity Calibration for Massive MIMO: Proposal, Modeling and Validation}

\author{
\IEEEauthorblockN{Joao Vieira, Fredrik Rusek, Ove Edfors, Steffen Malkowsky, Liang Liu, Fredrik Tufvesson} \\
\IEEEauthorblockA{Dept. of Electrical and Information Technology, Lund University, Sweden \\
firstname.lastname@eit.lth.se } }

\markboth{}%
{Shell \MakeLowercase{\textit{et al.}}: Bare Demo of IEEEtran.cls for Journals}

\maketitle

\begin{abstract}

This paper presents a mutual coupling based calibration method for time-division-duplex massive MIMO systems, which enables downlink precoding based on  uplink channel estimates. The entire calibration procedure is carried out solely at the base station (BS) side by sounding all BS antenna pairs. 
An Expectation-Maximization (EM) algorithm is derived, which processes the measured channels in order to estimate calibration coefficients. The EM algorithm outperforms  current state-of-the-art narrow-band calibration schemes in a mean squared error (MSE) and sum-rate capacity sense. Like its predecessors, the EM algorithm is general in the sense that it is not only suitable to calibrate a co-located massive MIMO BS, but also very suitable for calibrating multiple BSs in distributed MIMO systems.

The proposed  method is validated with experimental evidence obtained from a massive MIMO testbed. In addition, we address the estimated narrow-band calibration coefficients as a stochastic process across frequency, and study the subspace of this process based on measurement data. With the insights of this study, we propose an estimator which exploits the structure of the process in order to reduce the calibration error across frequency. A model for the calibration error is also proposed based on the asymptotic properties of the estimator, and is  validated with measurement results.

\end{abstract}


\section{Introduction }\label{sec:introduction}

\lettrine[lines=2]{M}{assive} Multiple-input Multiple-output (massive MIMO) is an emerging technology with the potential to be included in next generation wireless systems, such as fifth-generation (5G) cellular systems. Massive MIMO departs from traditional multi-user MIMO approaches by operating with a  large number of base station (BS) antennas, typically in the order of hundreds or even thousands, to serve a relatively small number of mobile terminals \cite{LUP2174140}. Such a system setup results in a multitude of BS antennas that can be used in an advantageous manner from multiple points of view \cite{LUP4305564}. 

One major challenge of operating with a large number BS antennas is that it renders explicit channel estimation in the downlink impractical. Basically, the  overhead of  channel estimation in the downlink and feeding back the channel estimate to the BS, scales linearly with the number of BS antennas, and quickly becomes unsupportable in mobile time-varying channels \cite{978730}. To deal with this challenge, the approach adopted is to operate in time-division-duplex (TDD) mode, rely on channel reciprocity, and use uplink channel state information (CSI) for downlink precoding purposes   \cite{Marzetta}. 
However, the presence of the analog front-end circuitry in practical radio units complicates the situation and makes the baseband-to-baseband channel non-reciprocal.
Explained briefly, the baseband representation of the received signals \cite{Proakis} experience channels that are not only determined by the propagation conditions, but also by the transceiver front-ends at both sides of the radio link. While it is generally agreed that the propagation channel is reciprocal \cite{Balanis}, the transceiver radio frequency (RF) chains  at both ends of the link are generally not \cite{5722383}. 
Hence, in order to make use of the reciprocity assumption and rely on the uplink CSI to compute precoding coefficients, the non-reciprocal transceiver responses need to be  calibrated. Such a procedure is often termed reciprocity calibration, and contains two steps: (i) estimation of calibration coefficients, and (ii) compensation by applying those to the uplink channel estimates.\footnote{ However, with the term reciprocity calibration, we will  interchangeably refer to the estimation step, compensation step, or both. The context will, hopefully, make clear which of the previous cases is being addressed.}

Reciprocity calibration of small scale TDD MIMO channels has been a matter of study in recent years. Depending on the system setup and requirements, the approach adopted can take many forms. For example, \cite{5722383} proposed a methodology based on bi-directional measurements between the two ends of a MIMO link to estimate suitable reciprocity calibration coefficients. This calibration approach falls in the class of "over-the-air"  calibration schemes where users are involved in the calibration process. A different approach is to rely on dedicated hardware circuitry for calibration purposes, see \cite{6573241,966592}. Despite the possibilities of extending both mentioned calibration approaches to a massive MIMO context, e.g., \cite{6963664,7416977}, recent calibration works suggest this is more difficult than previously thought. For example, \cite{7239634} questions the feasibility of having dedicated circuits for calibration when the number of transceivers to be calibrated grows large, and \cite{6760595} argues that the calibration protocols should preferably not rely on mobile units.  It thus appears that an increasing trend in massive MIMO systems is to carry out the calibration entirely at the BS side only through over-the-air measurements.

The first  proposal in this vein  was presented in \cite{ArgosThesis}. The work proposes an estimator for the calibration coefficients, which only makes use of channel measurements between BS antennas. More specifically, bi-directional channel measurements between a given BS antenna, so-called reference antenna, and all other antennas. This estimator was later generalized in order to calibrate large-scale distributed  MIMO networks \cite{6502966,6760595}. The estimation problem is formulated as  constrained least-squares (LS) problem where the objective function  uses channel measurements from a set of arbitrary antenna pairs of the network. The generality of this approach spurred many publications dealing with particular cases \cite{Wei2015,7032189,J7037384}. Parallel work in mutual coupling based calibration  was also conducted in \cite{7239634}. An estimator for the calibration coefficients, which enables maximum ratio transmission (MRT), was proposed for BS antenna arrays with  special properties.

Although it appears that over-the-air reciprocity calibration only involving the BS side is feasible, some matters need further investigation. Firstly, the approaches available in the literature for co-located BSs are not of great practical convenience. They either rely on antenna elements that need to be (carefully) placed in front of the BS antenna array solely for calibration purposes \cite{ArgosThesis}, or are only available for a restrictive case of antenna arrays \cite{7239634}. Secondly, most estimators for calibration  have been derived from empirical standpoints, e.g., \cite{7239634,ArgosThesis},  and respective extensions  \cite{6502966,7032189,J7037384}. 
It is not clear how far from fundamental estimation performance bounds, or how close to Maximum likelihood (ML) performance, such estimators are. Thirdly, most available calibration approaches are proposed for narrow-band systems. Such systems bandwidths are usually defined by the frequency selectivity of the propagation channel, which is typically much smaller than the frequency selectivity of the transceiver responses. This results in similar calibration coefficients for adjacent narrowband channels. Thus, it is of interest to model the statistical dependency of such calibration coefficients, and provide means to exploit this dependency in order to reduce the calibration error across frequency. Lastly, there is little publicly available  work on validation of massive MIMO calibration schemes. The need for validation is high, as it helps answering many questions of practical nature. For example, \cite{7340979} raises the question  whether the channel reciprocity assumption holds when strong coupling between BS antennas exist, and \cite{EURECOM+4463} questions if calibration assumptions similar to the ones used in this work, hold for massive MIMO arrays.


\subsection{Main Contributions of the Paper }

Below, we summarize the main contributions of this work.

\begin{itemize}
\item We propose a convenient calibration method mainly relying on mutual coupling between BS antennas to calibrate its non-reciprocal analog front-ends. We make no assumptions  other than channels due to mutual coupling  being reciprocal.

\item We show that the narrow-band calibration coefficients can be estimated by solving a joint penalized-ML estimation problem. We provide an asymptotically   efficient algorithm to compute the joint solution, which is a particular case of the EM algorithm.

\item We validate our calibration method experimentally  using a software-defined radio massive MIMO testbed. More specifically, we verify how the measured Error-Vector-Magnitude (EVM) of the downlink equalized signals decreases as the calibration accuracy increases, in a setup where three  closely spaced single-antenna users are spatially multiplexed by one hundred BS antennas.


\item We propose a non-white Gaussian model for the narrow-band calibration error based on the properties of the proposed estimator, and partially validate this model with measurements. 

\end{itemize}

\subsection{Notation}
\label{sec:NotationAndOutline}

The operators $(\cdot)^*$, $(\cdot)^{ T}$, $(\cdot)^{ H}$, and $(\cdot)^{ \dagger}$ denote  element-wise complex conjugate,  transpose,  Hermitian transpose, and  Moore-Penrose pseudo-inverse, respectively. The element in the $n$th row and $m$th column of matrix $\bf A$ is denoted by $\big[ {\bf A} \big]_{n,m}$. The operator $\mathrm{E}\left\{ \cdot \right\}$  denotes the  expected value.  $\operatorname{Re}\left\{\cdot\right\}$ and  $\operatorname{Im}\left\{\cdot\right\}$ return the real and imaginary part of their arguments. The matrix ${\bf I}$ denotes the identity matrix, and  ${\rm diag}\left\{a_1, a_2, \hdots a_M \right\} $ denotes an $M \times M$ diagonal matrix with diagonal entries given by $a_1, a_2, \hdots, a_M$. The operator $\ln$ denotes the natural logarithm. The set of the complex numbers and the set containing zero and the real positive numbers  are denoted by $\mathbb{C}$ and  $\mathbb{R}_{\geq 0}$, respectively. The operator $\setminus$ denotes the  relative set complement. Finally, $||\cdot||$ denotes the Frobenius norm.

\subsection{Paper Outline}

The remaining sections of the paper are as follows.  Section \ref{sec:ChannelModel} presents the  signal models.  Section \ref{sec:CREC} introduces the state-of-the-art estimator for the calibration coefficients, proposes a novel estimator, and provides a comparative analysis by means of   MSE and downlink sum-rate capacities. Section \ref{sec:Validation} validates the proposed calibration method experimentally. Using the estimated calibration coefficients obtained from the experiments, the purpose of Section \ref{sec:WB_EM} is twofold: \textit{i)} it studies several aspects of the calibration coefficients across $4.5$ MHz of transceiver bandwidth, \textit{ii)} it proposes a model for the calibration error of a narrowband system. Lastly, Section \ref{sec:conclusions} summarizes the key takeaways from this work.


\section{Signal Models }
\label{sec:ChannelModel}
This section starts by introducing the uplink and downlink signal models, and shows how downlink precoding can be performed using calibrated uplink channel estimates. Finally, it models the channels between BS antennas which we use for calibration purposes.

\subsection{Uplink and Downlink Signal models }


Let $K$ single-antenna users simultaneously transmit a pilot symbol in the uplink of a narrow-band MIMO system (e.g., a particular sub-carrier of an OFDM-MIMO system).  Collecting the pilot symbols in the vector ${\bf p} = [p_1 \cdots p_K]^{ T}$, the received signal by an  $M$-antenna base station can be written as
\begin{align}
\label{eq:UpLinkCh}
{\bf y}_{\rm UP} & =  {\bf H}_{\rm UP} \, {\bf p} + {\bf w}  \nonumber \\
 & =   {\bf R}_{\rm B} {\bf H}_{\rm P} {\bf T}_{\rm U} \, {\bf p}+ {\bf w}.
\end{align}
In (\ref{eq:UpLinkCh}), the  matrix ${\bf R}_{\rm B}~=~{\rm diag}\left\{\! r^{\rm B}_{1},\cdots, r^{\rm B}_{M} \!\right\}$ models the hardware response of $M$ BS receive RF chains (one RF chain per antenna), and the matrix  ${\bf T}_{\rm U}~=~{\rm diag}\left\{\! t^{\rm U}_{1},\cdots,t^{\rm U}_{K} \!\right\}$  models the hardware response of $K$ transmit RF chains (one chain per user).
${\bf H}_{\rm P}$ is the propagation channel matrix,
${\bf H}_{\rm UP}$ is the, so-called, uplink radio  channel, and ${\bf w}$ is a vector modeling uplink noise. Under the reciprocal assumption of the propagation channel, the received downlink signal can be written as
\begin{align}
\label{eq:DownLinkCh}
{\bf y}_{\rm DL}  = & {\bf H}_{\rm DL} \, {\bf z}{\rm '} + {\bf w}{\rm '}  \nonumber \\
= & {\bf R}_{\rm U}  {\bf H}_{\rm P}^{ T} {\bf T}_{\rm B} \, {\bf z}{\rm '}+ {\bf w}{\rm '}.
\end{align}
In (\ref{eq:DownLinkCh}), the  matrix ${\bf R}_{\rm U}~=~{\rm diag}\left\{\! r^{\rm U}_{1},\cdots, r^{\rm U}_{K} \!\right\}$ models the hardware response of the receive RF chains of the $K$ users, and the matrix  ${\bf T}_{\rm B}~=~{\rm diag}\left\{\! t^{\rm B}_{1},\cdots,t^{\rm B}_{M} \!\right\}$  models the hardware response of $M$ BS transmit RF chains. The entries of $\bf w{\rm '}$ model downlink noise, ${\bf H}_{\rm DL}$ is the downlink radio channel, and ${\bf z}{\rm '}$ is a vector with linearly precoded QAM symbols. In particular,  ${\bf z}{\rm '}=\bf Px$, where $\bf P$ is the precoding matrix, and the entries of $\bf  x$ contain QAM symbols.

\subsection{Calibration Coefficients }

Assume that an error free version of the uplink radio channel, ${\bf H}_{\rm UP}$, is available at the BS. The transpose of the result of pre-multiplying ${\bf H}_{\rm UP}$ with the matrix  $\alpha {\bf T}_{\rm B}{\bf R}_{\rm B}^{-1}$, where $\alpha \in \mathbb{C}\setminus 0$ and $r_m \neq 0, \forall \; m$, is a matrix ${\bf G}$ that, if used for precoding purposes by means of a linear filtering, is sufficient for spatially multiplexing terminals in the downlink with reduced crosstalk. This can be visualized by expanding ${\bf G}$ as
\begin{align}
{\bf G}  =  &  \left( \left( \alpha {\bf T}_{\rm B}{\bf R}_{\rm B}^{-1}  \right) { \bf H}_{\rm UP} \right)^{ T} \nonumber \\
= & \; \alpha {\bf T}_{\rm U} {\bf H}_{\rm P}^{ T}  {\bf T}_{\rm B}  \nonumber  \\
\label{eq:lastEq}
= & \;   \alpha {\bf T}_{\rm U} {\bf R}_{\rm U}^{-1} {\bf H}_{\rm DL}.
\end{align}
From (\ref{eq:lastEq}) we have that ${\bf G}$ is effectively the \textit{true} downlink radio channel ${\bf H}_{\rm DL}$ pre-multiplied with a diagonal matrix with unknown entries accounting for the  user terminals responses $ {\bf T}_{\rm U} {\bf R}_{\rm U}^{-1}$, and $\alpha$. The row space of ${\bf G}$ is thus the same as of the downlink radio channel ${\bf H}_{\rm DL}$. This is a sufficient condition to cancel inter-user interference if, for example, ZF precoding is used (i.e., ${\bf H}_{\rm DL}  {\bf G}^\dagger $ is a diagonal matrix). 

From  (\ref{eq:lastEq}), it can also be seen that any non-zero complex scalar $\alpha$ provides equally good calibration.\footnote{This follows since both magnitude and phase of $\alpha$ are not relevant in this calibration setup. The former holds since any real scaled channel estimate provides the same precoder matrix $\bf P$, if the precoder has a fixed norm. The latter follows from  (\ref{eq:lastEq}), since the (uniform phases of the) diagonal entries of $ {\bf T}_{\rm U} {\bf R}_{\rm U}^{-1}$ are unknown  to the precoder in this calibration setup.} Thus, the matrix
\begin{align}
{\bf C} = & { \rm diag\{c_1,\cdots,c_M \}} \nonumber \\
= &  {\bf T}_{\rm B}{\bf R}_{\rm B}^{-1} 
\end{align}
is the, so-called, calibration matrix, and $\left\{c_m\right\}$ are the calibration coefficients which can be estimated up to a common complex scalar $\alpha$. We remark that, although not strictly necessary to build estimators, the concept of a reference transceiver \cite{ArgosThesis} can be used to deal with the ambiguity of estimating $\left\{c_m\right\}$ up to $\alpha$.\footnote{Explained briefly, assuming $c_\textit{ref}=1$ and solving for $\left\{c_m\right\}\setminus c_\textit{ref}$, where $c_\textit{ref}$ is the calibration coefficient associated with a reference transceiver. } 
The remainder of the paper deals with estimation aspects of $c_m= t^{\rm B}_m/r^{\rm B}_m$. Thus, for notational simplicity, we write $t_m=t^{\rm B}_m$, $r_m=r^{\rm B}_m$, ${\bf R} = {\bf R}_{\rm B}$, and ${\bf T} = {\bf T}_{\rm B}$. Also, we stack $\left\{c_m\right\}$ in the vector ${\bf c} = [c_1 \cdots c_M]^{ T}$, for later use.



\subsection{Inter-BS Antennas Signal model }

To estimate the calibration coefficients $c_m$ we sound the $M$ antennas one-by-one by transmitting a sounding signal from each one and receiving on the other $M-1$ silent antennas. Let the sounding signal transmitted by antenna $m$ be $s_m = 1, \forall \; m$,  unless explicitly said otherwise. Also, let $y_{n,m}$ denote the signal received at antenna $n$ when transmitting at antenna $m$. It follows that the received signals between any pair of antennas can be written as
\begin{equation}
\label{eq:SystemModel33}
\left[ \begin{array}{cc} y_{n,m}\\ y_{m,n}  \end{array} \right] =    h_{n,m} \left[ \begin{array}{cc} r_{n} t_{m} & 0  \\ 0 & r_{m} t_{n}  \end{array} \right]  \left[ \begin{array}{cc} s_{m}  \\ s_{n}  \end{array} \right] + \left[ \begin{array}{cc} n_{n,m} \\ n_{m,n}   \end{array} \right],
\end{equation}
where
\begin{align}
\label{eq:CouplingMOdel}
  h_{n,m}  = & \; \bar{h}_{n,m} + \tilde{h}_{n,m} \\
           = & \; |\bar{h}_{n,m}| \exp( j2\pi\phi_{n,m} ) + \tilde{h}_{n,m}
\;  
\end{align}
models the (reciprocal) channels between BS antennas. The first term $\bar{h}_{n,m}$ describes a channel component due to mutual coupling between antenna elements, often stronger for closely spaced antennas, which we lay down a model for in Sec. \ref{sec:ModMutualCoupling}. The terms $|\bar{h}_{n,m}|$ and $\phi_{n,m}$ denote the magnitude and phase of $\bar{h}_{n,m}$, respectively. The term $\tilde{h}_{n,m}$, which absorbs all other channel multipath contributions except for the mutual coupling  (e.g., reflections by scatterers in front of the BS) is modeled  by an i.i.d. zero-mean circularly symmetric complex Gaussian random variable with  variance $\sigma^2$.
Non-reciprocal channel components are modeled by $r_m$ and $t_m$ which materially map to the cascade of hardware components, mainly in the analog front-end stage of the receiver and transmitter, respectively. We assume i.i.d. circularly symmetric zero-mean complex Gaussian noise contributions $n_{m,n}$  with  variance $N_0$. Letting $\big[ {\bf Y} \big]_{m,n} = y_{m,n}$, the received signals can be expressed more compactly as
\begin{equation}
\label{eq:SignalModel}
\bf Y = R H T + N.
\end{equation}
Note that ${\bf H} = {\bf H}^{ T}$ is assumed, and the diagonal entries in the $M \times M$ matrix $\bf Y$ are undefined. 

\subsection{Modeling Mutual Coupling}
\label{sec:ModMutualCoupling}

The purpose of this section is to provide a model for the mutual coupling  between antenna elements, i.e. $\bar{h}_{m,n}$, as a function of their distance. 
Instead of pursuing a circuit theory based approach to model the effect of mutual coupling \cite{7340979}, our modeling approach uses S-parameter measurements from a massive MIMO BS antenna array \cite{1142529}.
 We note that this model is used only for simulation purposes, and not to derive any of the upcoming estimators of $\bf c$.


\subsubsection{Test Array Description}
\label{sec:TestArray}
\begin{figure}[t]
    \centering
    \includegraphics[scale=.067]{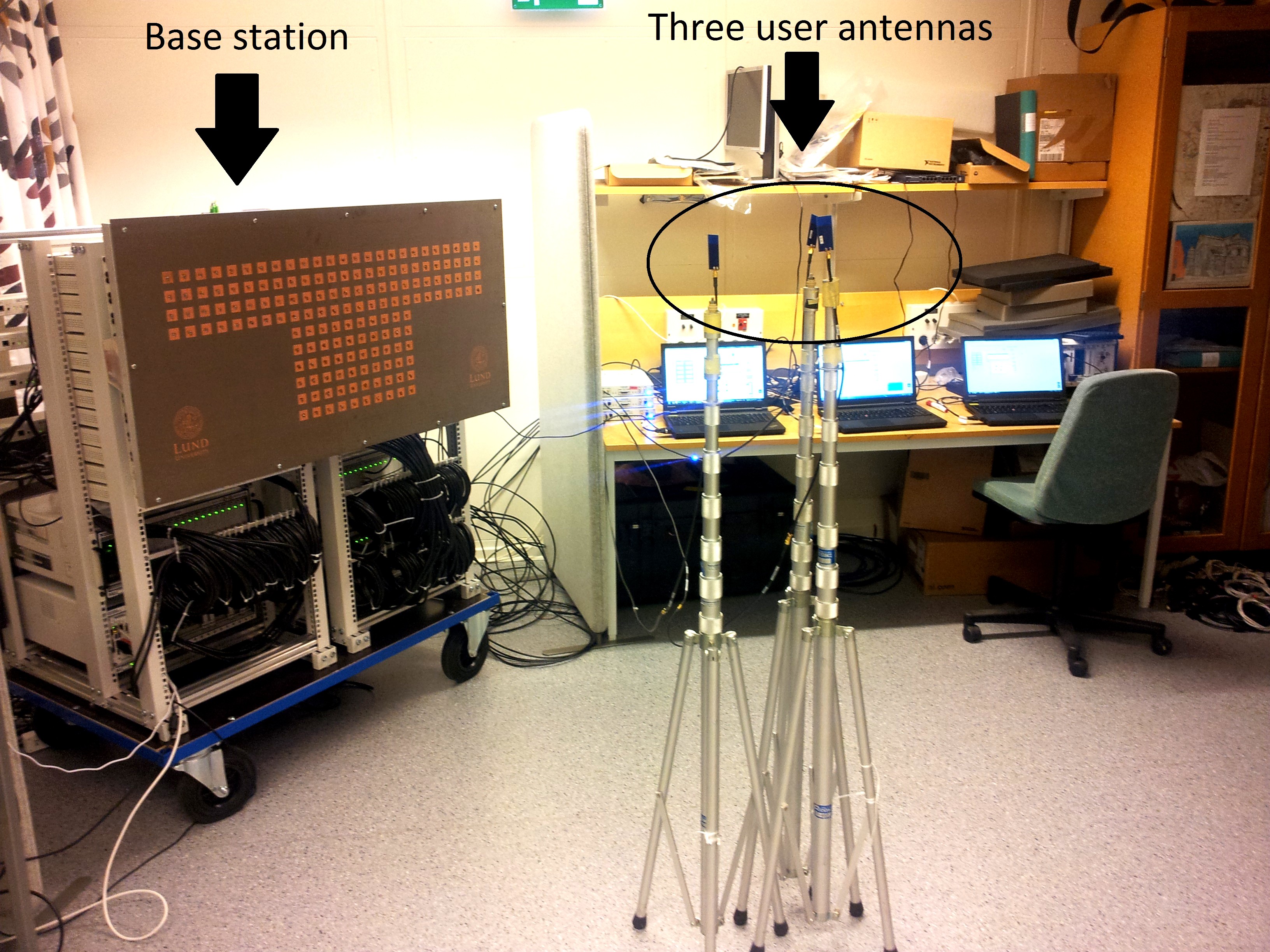}
    \caption{The massive MIMO lab setup used throughout this work. The BS is on the left side where a "T" shaped antenna array can be seen. Three closely spaced user antennas stand the middle of the picture.}
    \label{fig:DownlinkSetup}
\end{figure}

The antenna array considered for modeling is a 2-dimensional planar structure with dual-polarized patch elements spaced by half a wavelength. More information about the antenna array can be found in  \cite{BasestationPaper}. The dimensional layout of the array adopted for this work corresponds to the $4 \times 25$ rectangular grid in the upper part of the array shown in Fig. \ref{fig:DownlinkSetup}. Only one antenna port  is used per antenna element. 
For a given antenna, the polarization port is chosen such that its adjacent antennas - the antennas spaced by half wavelength - are cross-polarized. This setting provides, so-called, polarization diversity, and reduces mutual coupling effects between adjacent antennas since co-polarized antennas couple stronger \cite{1142529}.

\subsubsection{Modeling coupling gains between antennas}
\label{subsec:Modelingcoup}

\begin{figure}[t]
    \centering
    \includegraphics[scale=.73]{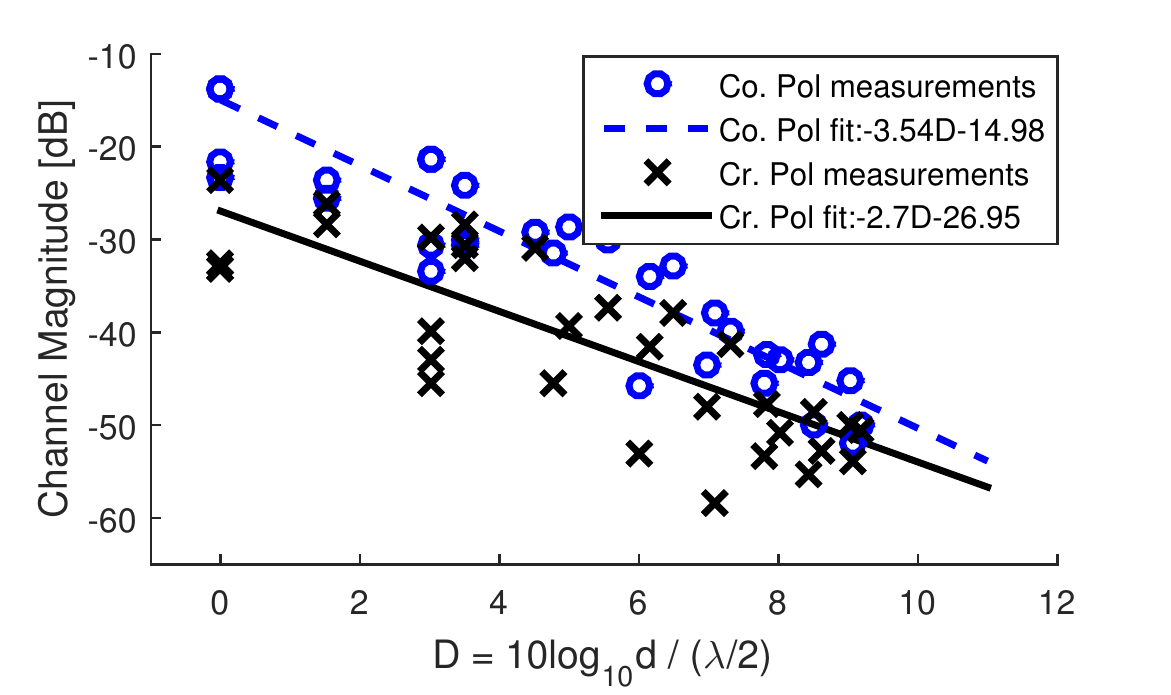}
    \caption{Measured coupling magnitudes $|\bar{h}_{n,m}|$ between different antenna pairs. The circles corresponds to measurements between co-polarized antenna elements, and the crosses between cross polarized antenna elements. The variable $\rm d$ corresponds to the physical distance between antenna elements. The straight lines represent the corresponding linear LS fits. }
    \label{fig:CouplingModelling}
\end{figure}

The channel magnitude $|\bar{h}_{n,m}|$ between several pairs of cross and co-polarized antennas were measured in an anechoic chamber using a Vector Network Analyzer, at 3.7 GHz -  the center frequency of the array. 
Fig. \ref{fig:CouplingModelling} shows the measured channel magnitudes.  Different channel magnitudes for the very same measured distance and polarization cases, are due mostly to the relative orientation of the antenna pair with respect to their polarization setup. For example, vertically (co-)polarized antennas couple more strongly when they are oriented horizontally.
A linear LS fit was performed to model the coupling gain $|\bar{h}_{n,m}|$ as a function of antenna distance. 
The phase $\phi_{m,n}=\phi_{n,m}$ is modeled uniformly in $[0,1]$, as a clear dependence with distance was not found.

\section{Estimation of the Calibration Coefficients} \label{sec:CREC}

In this section we deal with estimation aspects of the calibration  matrix ${\bf C=TR}^{-1}$. 
We introduce the state-of-art estimator of $\bf C$ \cite{6502966,6760595}, and propose a novel iterative penalized-ML estimator.\footnote{We note that the only assumption used to derive the estimators is  ${\bf H} = {\bf H}^{ T}$. The generality of this assumption allows the estimators to be used in other calibration setups than those of co-located MIMO systems, as it will be pointed out later.} A comparative numerical analysis is made by means of MSE and  sum-rate capacity.  We conclude the section with two interesting remarks.

\subsection{The Generalized Method of Moments estimator  }
\label{sec:GMMe}

Calibration of large-scale distributed MIMO systems using a similar system model to (\ref{eq:SignalModel}) was performed in \cite{6760595} and \cite{6502966}.\footnote{In their work,  $h_{m,n}$ denotes the propagation channel between antennas of different BSs. The reciprocal model adopted for $h_{m,n}$ accounts for large-scale and small-scale fading.}  Based on the structure of the system model, the authors identified that 
\begin{equation}
\label{eq:MomentCondition}
 \mathrm{E} \left\{ y_{n,m}c_n - y_{m,n}c_m  \right\}  = 0 . 
\end{equation}
Define $ g_{m,n} \triangleq  y_{n,m}c_n - y_{m,n}c_m $, and  ${ \boldsymbol{g}({\bf c}) }= \left[ g_{1,2} \dots g_{1,M} \, g_{2,3} \dots g_{2,M} \dots g_{M-1,M} \right] ^T$.\footnote{The dependency of $ \boldsymbol{g}({\bf c})$ on $y_{n,m}$ is explicitly left out, for notational convenience.} An estimator for $\bf c$ was proposed by solving
\begin{equation}
\label{eq:GMMeCost}
\hat{\boldsymbol{c}}_{\rm GMM} =  \arg \min_{ \substack{\boldsymbol{c} \\ s.t.\; f_c(\boldsymbol{c})=1 }} {  \boldsymbol{g}^{ H}({\bf c}) \boldsymbol{W} \boldsymbol{g}({\bf c}})
\end{equation}
with  $\bf \boldsymbol{W} = \boldsymbol{I}$.  Two constraints   were suggested to avoid the all-zero solution, namely $f_c(\boldsymbol{c})=c_1$ or $f_c(\boldsymbol{c})=||{\bf c} ||^2$. By setting the gradient  with respect to ${\bf c}$ to zero, an estimator in closed-form was given. Next, we provide a few remarks on this estimation approach.

A fact not identified in  \cite{6760595} and \cite{6502966}, is that this estimator is an instance of a estimation framework widely used for statistical inference in econometrics, namely the generalized method of moments (GMM). The variable $g_{m,n}$ - whose expectation is zero - is termed a moment condition within  GMM literature \cite{hall2004generalized}.  With a proper setting of the weighting matrix $\bf \boldsymbol{W}$, it can be shown that the solution to ($\ref{eq:GMMeCost}$) provides an estimator that is asymptotically efficient\cite{hall2004generalized}. However, no such claim can be made in the low signal-to-noise (SNR) regime, where an optimal form of $\bf \boldsymbol{W}$ is not available in the literature. This typically leads to empirical settings of $\bf \boldsymbol{W}$, e.g., $\bf \boldsymbol{W} = \boldsymbol{I}$. As a result, moment conditions comprising measurements with low SNR    constrain the performance  since they are weighted equally. It thus appears that an inherent problem of the GMM estimator is the selection of $\bf \boldsymbol{W}$. Nevertheless, it provides a closed-form estimator based on a cost function  where nuisance parameters for calibration, as $h_{m,n}$, are conveniently left out.

\subsection{Joint Maximum Penalized-Likelihood estimation}

Here we address  joint maximum penalized-likelihood estimation for  $\bf c$ and for the equivalent channel $\bf \Psi \triangleq RHR$.  Noting that (\ref{eq:SignalModel}) can be written as 
\begin{align}
\bf Y = &  \; \bf R H R C + N \nonumber \\
      = &  \; \bf \Psi C + N, 
      \label{eq:EqModel}
\end{align}
the optimization problem  can be put as
\begin{align}
\label{eq:JointLikeIni}
\bf [ \hat{c}, \; \hat{\Psi}] & =  \arg \max_{ \bf c, \Psi }  \ln  p( {\bf Y | C, \Psi } )  +  {\rm Pen}({\bf C,\Psi},\epsilon')   \nonumber \\
 & = \arg \min_{ {\bf c}, \Psi } J_{\rm ML}( {\bf Y, C, \Psi},\epsilon)
\end{align}
with 
$
J_{\rm ML}( {\bf Y, C, \Psi},\epsilon) =  || {\bf Y - \Psi  C } ||^2 +  {\rm Pen}({\bf C,\Psi},\epsilon).
$
Here, $p( {\bf Y | C, \Psi })$ denotes the probability density function (PDF) of $\bf Y$ conditioned on $\bf C$ and $\bf \Psi$, and ${\rm Pen}({\bf C,\Psi},\epsilon)$ is a penalty term parametrized by $\epsilon=\epsilon' N_0$ with $\epsilon \in \mathbb{R}_{\geq 0}$. 

There are many uses for the penalty term in ML formulations \cite{Bishop}.  Here, we use it mainly to control the convergence rate of the algorithm (presented in Sec. \ref{subsec:EMalg}), and use $\epsilon$ as a tuning parameter. 
With this in mind, we pursue Ridge Regression  and set the penalty term as\footnote{Ridge Regression \cite{Hoerl} is an empirical regression approach  widely used in many practical fields, e.g., Machine Learning \cite{Bishop}, as it provides  estimation robustness when the model is subject to a number of degeneracies. This turns out to the case in this work, and we point out why this occurs later. However, we emphasize that the main reason of adding the penalty terms is to control the convergence of the algorithm, which we also point out later why this is the case. To finalize, we parametrize the penalty term (\ref{eq:Penaltyterm}) with a single  parameter  in order simplify the convergence analysis and be able to extract meaningful insights.}
\begin{equation}
\label{eq:Penaltyterm}
{\rm Pen}({\bf C,\Psi},\epsilon) = \epsilon ( || {\bf C}||^2 + ||{\bf \Psi}||^2 ).
\end{equation}

After some re-modeling, a vectorized version of (\ref{eq:EqModel}) can be written as
\begin{equation}
\label{eq:EqModel10}
\bf  \widetilde{Y}  = \bf \Psi_{\rm eq}({\bf \tilde{\Psi}}) c +  \widetilde{N},
\end{equation}
or as
\begin{equation}
\label{eq:EqModel11}
\bf {\bf Y'} = \bf C_{\rm eq}({\bf c })  {\bf \tilde{\Psi}} + N',
\end{equation}
where ${\bf \tilde{\Psi}}$ stacks all $\psi_{n,m}=[{\bf \Psi}]_{n,m}$ into an $ (M^2-M)/2 \times 1$ vector,  and $\bf \Psi_{\rm eq}({\bf \tilde{\Psi}})$ and $\bf C_{\rm eq}({\bf c })$ are equivalent observation matrices which are constructed from ${\bf \tilde{\Psi}}$ and  $\bf c$, respectively. The structure of these matrices is shown in Appendix A, but it can be pointed out that $\bf \Psi_{\rm eq}({\bf \tilde{\Psi}})$ and $\bf C_{\rm eq}({\bf c })$ are a block diagonal, where each block is a column vector. 


From (\ref{eq:EqModel11}), it is seen that for a given $\bf \bf C_{\rm eq}({\bf c })$, the penalized-ML estimator of ${\bf \tilde{\Psi}}$ is given by\footnote{The factor $2$ in the regularization term of (\ref{eq:LS}) appears since $\psi_{m,n}=\psi_{n,m}$. Note that  $\epsilon$ is considered as a constant during the optimization, otherwise it is obvious that  $\epsilon=0$ minimizes (\ref{eq:Penaltyterm}).}
\begin{equation}
\label{eq:LS}
{\bf \tilde{\Psi}}_{\rm ML}  =  \Big( {\bf C}_{\rm eq}^{ H}({\bf c}) {\bf C_{\rm eq}(\bf c)} + 2 \epsilon  {\bf I} \Big)^{-1}  {\bf C}_{\rm eq}^{ H}(\bf c){\bf Y'},
\end{equation}
 If in (\ref{eq:EqModel11}), we replace ${\bf \tilde{\Psi}}$ by its  estimate ${\bf \tilde{\Psi}}_{\rm ML}$, then the penalized ML solution for $\bf c$ is 
 \iftoggle{2Collumn}{%
\begin{align}  
\label{eq:EqModel5}
{\hat{{\bf c}}_{\rm ML}} = \arg \min_{ {\bf c} }   ||  { \bf Y' - \bf C_{\rm eq}({\bf c })}&\Big( {\bf C}_{\rm eq}^{ H}({\bf c}) {\bf C_{\rm eq}(\bf c)} + 2 \epsilon  {\bf I} \Big)^{-1} \nonumber \\ 
& \times {\bf C}_{\rm eq}^{ H}({\bf c}){\bf Y'}  ||^2,
\end{align}
}{
\begin{equation}  
\label{eq:EqModel5}
{\hat{{\bf c}}_{\rm ML}}=\arg \min_{ {\bf c} }   ||{ \bf Y' - \bf C_{\rm eq}({\bf c })}\Big( {\bf C}_{\rm eq}^{ H}({\bf c}) {\bf C_{\rm eq}(\bf c)} + 2 \epsilon  {\bf I} \Big)^{-1}  {\bf C}_{\rm eq}^{ H}({\bf c}){\bf Y'}  ||^2,
\end{equation}
}
It is possible to further simplify $(\ref{eq:EqModel5})$ for the case of unpenalized ML estimation ($\epsilon =0$) and attack the optimization problem with gradient-based methods \cite{scharf1991statistical}. We have implemented the conjugate gradient method in a Fletcher-Reeves setting with an optimized step-size through a line-search. However, this turns out to be far less robust than, and computationally more expensive to, the method provided next. Therefore we omit to provide the gradient in closed form.

%

\subsection{An EM Algorithm to find the joint Penalized-ML Estimate}
\label{subsec:EMalg}
Here we provide a robust and computational efficient algorithm to find the joint penalized-ML estimate of $\bf c$ and $\bf \Psi$. Instead of pursuing an approach similar to the one used to reach (\ref{eq:EqModel5}), the algorithm  has its roots in the joint solution found by setting the gradient of  $ J_{\rm ML}( {\bf Y, C, \Psi,\epsilon})$ to zero. Before presenting the algorithm, we therefore briefly address this gradient approach. 

Each entry of ($\ref{eq:EqModel}$) is given by  $y_{n,m} = \psi_{n,m}c_m + n_{n,m}$. The derivative of $J_{\rm ML}( {\bf Y, C, \Psi,\epsilon})$ with respect to $c_m^*$ is given by
\begin{equation}
\label{eq:Grad}
\frac{ \partial J_{\rm ML}( {\bf Y, C, \Psi,\epsilon})   } { \partial c_m^* } = \epsilon  c_m +  \sum\limits_{\substack{n =1 \\ n\neq m}}^{M} |\psi_{n,m}|^2  c_m - y_{n,m}\psi^*_{n,m}.   
\end{equation}
Setting (\ref{eq:Grad}) to zero and solving for $c_m$ yields
\begin{equation}
\label{eq:CmInScalars}
c_m  = \left(  \epsilon +  \sum\limits_{\substack{n =1 \\ n\neq m}}^{M}  |\psi_{n,m}|^2  \right)^{-1} \sum\limits_{\substack{n =1 \\ n\neq m}}^{M} \psi^*_{n,m} y_{n,m} , 
\end{equation}
which can be expressed  in a vector form as
\begin{equation}
\label{eq:LSCss}
\hat{{\bf c}}_{\rm ML} =  \Big( {\bf \Psi}^{ H}_{\rm eq}({\bf \tilde{\Psi}} ) {\bf \Psi}_{\rm eq}({\bf \tilde{\Psi}} ) + \epsilon  {\bf I} \Big)^{-1} {\bf \Psi}^{ H}_{\rm eq}({\bf \tilde{\Psi}} ) { \bf \widetilde{Y}}.
\end{equation}
In a similar fashion, setting the derivative of  $J_{\rm ML}( {\bf Y, C, \Psi,\epsilon})$ with respect to $\psi_{n,m}^*$ to zero and solving for $\psi_{n,m}$ provides
\begin{equation}
\label{eq:PsiInScalars}
 \psi_{n,m}  = \left(  |c_{n}|^2 + |c_{m}|^2 +2 \epsilon \right)^{-1}  \left(  y_{m,n} c^*_{n} + y_{n,m} c^*_{m} \right),
\end{equation}
which can be expressed  in a vector form as (\ref{eq:LS}). Equations (\ref{eq:CmInScalars}) and (\ref{eq:PsiInScalars}) show the analytical form for each entry of the penalized-ML vector estimates, which will prove to be useful during the complexity analysis. Combining the results from (\ref{eq:LSCss}) and (\ref{eq:LS}) yield the joint  solution
\begin{equation}
\label{eq:JointMLsol}
\begin{bmatrix}
{ \hat{ \bf c }}_{\rm ML}      \\
{\bf \tilde{\Psi}}_{\rm ML}  
\end{bmatrix}
=
\begin{bmatrix}
\Big( {\bf \Psi}^{ H}_{\rm eq}({\bf {\bf \tilde{\Psi}}_{\rm ML}} ) {\bf \Psi}_{\rm eq}({\bf {\bf \tilde{\Psi}}_{\rm ML}} ) + \epsilon  {\bf I} \Big)^{-1} {\bf \Psi}^{ H}_{\rm eq}({\bf {\bf \tilde{\Psi}}_{\rm ML}} ) { \bf \widetilde{Y}} \\
\Big( {\bf C}_{\rm eq}^{ H}({\hat{{\bf c}}_{\rm ML} }) {\bf C_{\rm eq}({\hat{{\bf c}}_{\rm ML} })} + 2 \epsilon  {\bf I} \Big)^{-1}  {\bf C}_{\rm eq}^{ H}({\hat{{\bf c}}_{\rm ML} }){\bf Y'}
\end{bmatrix}
\end{equation}

The particular structure of (\ref{eq:JointMLsol}) suggests that a pragmatic approach for solving can be pursued. More specifically,  (\ref{eq:JointMLsol}) can be separated into two sub-problems, i.e., solving for $ { \hat{ {\bf c}}}_{\rm ML}  $ and ${\bf \tilde{\Psi}}_{\rm ML}$ separately. Since each of the solutions depend on previous estimates, the joint solution can be computed iteratively, by sequentially solving two separate regularized LS problems, given an initial guess. Since each iteration estimates $\bf c$ and  ${\bf \tilde{\Psi}}$ separately, this approach can be seen as an instance of the
 EM algorithm \cite{Kay}, where the - often challenging - \textit{Expectation step} is performed by estimating only the first moment of the nuisance parameters $\left\{\psi_{m,n}\right\}$. 
The convergence of the algorithm can be analyzed using standard methods, such as a distance between consecutive point estimates. 
The GMM estimator can be used to compute a reliable initial guess for iteration - in contrast to a purely random initialization. This is often good practice to ensure convergence to a \textit{suitable} local optimum since $J_{\rm ML}( {\bf Y, C, \Psi},\epsilon)$ is not a convex function of its joint parameter space.
For sake of clarity, Algorithm 1 summarizes the proposed iterative procedure.

Observe that $\epsilon$, i.e. the penalty term parameter in (\ref{eq:Penaltyterm}), ends up regularizing both matrix inversions. This is of notable importance from two points-of-view: \textit{i)} from an estimation (robustness) point-of-view, since the matrices to be inverted are constructed from parameter estimates (and thus are subject to estimation errors) and no favorable guarantee exists on their condition number, e.g., see (\ref{eq:CeqStruct}). \textit{ii)} from a convergence point-of-view, as it is well-known that the convergence rate of regularized LS adaptive filters is inversely proportional to their eigenvalue spread \cite{Haykin}; This property combo justifies why  Ridge Regression was pursued in the first place.


\begin{algorithm}[t]
    \caption{ Expectation-Maximization }
        \small
    \begin{algorithmic}[1]
        \REQUIRE Measurement matrix $\bf Y$, convergence threshold $\Delta_{\rm ML}$, penalty parameter $\epsilon$, initial guess  $\hat{\bf c}$
        \STATE \textbf{Initialization}:  set $\Delta  = \delta$ where $\delta > \Delta_{\rm ML} $
        \WHILE{$ \Delta \geq \Delta_{\rm ML}  $}
        \STATE  ${\bf \tilde{\Psi}}_{\rm ML}  = \Big( {\bf C}_{\rm eq}^{ H}({\bf \hat{c}}) {\bf C_{\rm eq}(\bf \hat{c})} + 2 \epsilon  {\bf I} \Big)^{-1}  {\bf C}_{\rm eq}^{ H}(\bf \hat{c}) { \bf Y'}$
        \STATE $\hat{{\bf c}}_{\rm ML} =  \Big( {\bf \Psi}^{ H}_{\rm eq}({\bf \tilde{\Psi}}_{\rm ML} ) {\bf \Psi}_{\rm eq}({\bf \tilde{\Psi}}_{\rm ML} ) + \epsilon  {\bf I} \Big)^{-1} {\bf \Psi}^{ H}_{\rm eq}({\bf \tilde{\Psi}}_{\rm ML} ) { \bf \widetilde{Y}}$ 
        \STATE $\Delta  =  || \hat{{\bf c}}_{\rm ML} -  \hat{{\bf c}} ||^2$
        \STATE $\bf \hat{c}=\hat{{\bf c}}_{\rm ML}$   
        \ENDWHILE
        \OUTPUT Calibration coefficients estimate $\hat{{\bf c}}_{\rm ML}$
    \end{algorithmic}
    \normalsize
    \label{table:iterativeML}
\end{algorithm}


A side remark regarding an application of the EM algorithm follows. We highlight that the calibration coefficients $\bf c$ and the equivalent channels  $\psi_{m,n}=r_m h_{m,n} r_n$ are jointly estimated. As previously mentioned, this a feature is not present in the GMM estimator. Noticeably, this feature makes the EM algorithm robust and hence very suitable to calibrate distributed MIMO systems since channel fading (i.e., high variations of $|h_{m,n}|$) often occurs \cite{6760595}. As mentioned in Sec. \ref{sec:GMMe},  the system model used can be also representative to that of distributed systems.



\subsection{Complexity Analysis }
\label{subsec:Complexity}


The complexity of each iteration of Algorithm  \ref{table:iterativeML} is dominated by steps 3 and 4.  Fortunately the block diagonal structure of the equivalent matrices allows for the inversions to be of reduced complexity, as detailed next. From (\ref{eq:PsiInScalars}), each calculation of  $\psi_{m,n}$  requires a few multiplications and additions. Since $\big(M^2 - M\big)/2$ such calculations are needed to compute (\ref{eq:LS}), the complexity order of step 3 is $\mathcal{O}(M^2)$. Similarly, the complexity  of step 4 is $\mathcal{O}(M^2)$ which can be seen directly from (\ref{eq:CmInScalars}). The explanation of the $\mathcal{O}(M^2)$ behavior is that the complexity of each calibration coefficient $c_m$ is $\mathcal{O}(M)$, and $M$ such calibration coefficients need to be computed. Overall, each iteration of the EM algorithm is of complexity  $\mathcal{O}(M^2)$, and the algorithm's complexity is $\mathcal{O}(N_{\rm ite} \, M^2)$, with $N_{\rm ite}$ being the number of iterations needed for convergence. The number of iterations needed for convergence is studied in  Sec. \ref{subsubsec:ConvEM}.


As for the GMM estimator, the closed-form solutions presented in \cite{6760595} and \cite{6502966}  have complexity orders of $\mathcal{O}(M^3)$, as they consist of an inverse of a Hermitian matrix of size $M-1$, and of the eigenvector associated with the smallest eigenvalue of a Hermitian matrix of size $M$.

On a practical note, we remark that the computational complexity of both approaches does not stand as a prohibitive factor for BS arrays using hundreds or even several thousands of antennas. This is because calibration typically needs to be performed on a hourly basis \cite{ArgosThesis,BasestationPaper}.



\subsection{Performance Assessment}


\subsubsection{Simulation setup for the MSE analysis}
\label{Sec:SimSetup}

We simulate reciprocity calibration over a $4 \times 25$ rectangular array as the one in Fig. \ref{fig:DownlinkSetup}. The linear regression parameters obtained in Sec. \ref{subsec:Modelingcoup} are used to model the coupling gains $\bar{h}_{m,n}$. The $m$th transceiver maps to the antenna in row $a_{\rm row}$ and column $a_{\rm col}$ of the  array as $m= 25 (a_{\rm row}-1) + a_{\rm col}$. The reference transceiver index is set to $\textit{ref}=38$, as it is associated with one of the most central antenna elements of the 2-D array.



The Cram\'{e}r-Rao Lower Bound (CRLB) is computed to verify the asymptotical properties of the estimators' error \cite{Kay}. From (\ref{eq:CouplingMOdel}) and (\ref{eq:SignalModel}), it can be seen that if $\bar{h}_{m,n}$ is assumed to be known, the PDF of $\bf Y$ conditioned on $\bf R$ and $\bf T$ is a multivariate Gaussian PDF. This makes the  CRLB of $\bf c$ to have a well known closed-form, which is computed in Appendix B.

The transmitter $t_m$  and receiver $r_m$ gains are set to $ t_m = ( 0.9 + \frac{0.2 m}{M} \exp(-j2 \pi m/M) ) /  t_\textit{ref}$ and $ r_m = ( 0.9 + \frac{0.2 (M-m)}{M} \exp(j2 \pi m/M) ) / r_\textit{ref}$, respectively. We used this deterministic setting for the transceivers, as it allows for a direct comparison of the parameter estimates' MSE with the CRLB. Moreover, this  setting incorporates eventual mismatches within the transceivers complex amplitude which are in line with the magnitude variations measured from the  transmitters/receivers of our testbed, i.e., spread of around 10-percent around the mean magnitude (and uniform phase). This spread is in line with transceiver models adopted in other calibration works \cite{6760595}.

The variance $\sigma^2$ of the multipath propagation contribution during calibration is set to $-60$ dB. Our motivation for this value is as follows. If the closest physical scatter to the BS is situated, say, 15 meters away, then by Friis' law \cite{molisch2010wireless} we  have a path loss of around $ 10 \log_{10}( \frac{4 \pi d }{\lambda} ) = 10 \log_{10}( \frac{4 \pi (2\times 15m) }{ 3 \times 10^8/(3.7 \times 10^9) } ) = 73$ dB per path. This number does not account for further losses due to reflections and scattering. Based on this, we use $-60$ dB as the power (variance) of the resulting channel stemming from a large number of such uncorrelated paths.

For consistency with the reference antenna concept used in the CRLB computations, the MSE of the EM algorithm output $\hat{{\bf c}}_{\rm ML}$, is defined as
\begin{align}
    \label{eq:MSENewDef}
    \mathrm{MSE}_m & =   \mathrm{E} \left\{    | c_m - \left[ \hat{{\bf c}}_{\rm ML} \right]_{ m,1}  / \left[ \hat{{\bf c}}_{\rm ML}\right]_{ \textit{ref},1} |^2 \right\},
\end{align}
since the estimated "reference" coefficient $\left[ \hat{{\bf c}}_{\rm ML}\right]_{ \textit{ref},1}$ is not necessarily equal to $1$. This is because the concept of reference antenna is not used by the EM algorithm. As for the GMM estimator, the constraint provided in \cite{6502966} is adopted, i.e., $c_\textit{ref}=1$ in (\ref{eq:GMMeCost}), which is already coherent with the computed CRLB. The results are averaged over 1000 Monte-Carlo simulations, and the threshold $\Delta_{\rm ML}$ is  set  to $10^{-6}$ which, based on our experience, ensures that convergence is reached in many parameter settings. The initial guess for the EM algorithm is produced by the GMM estimator.



%

\subsubsection{Estimators' MSE vs CRLB}
\label{subsubsec:MSEvsCRLB}
\begin{figure*}
    \center
    \begin{tabular}{cc}
        \includegraphics[scale=.77]{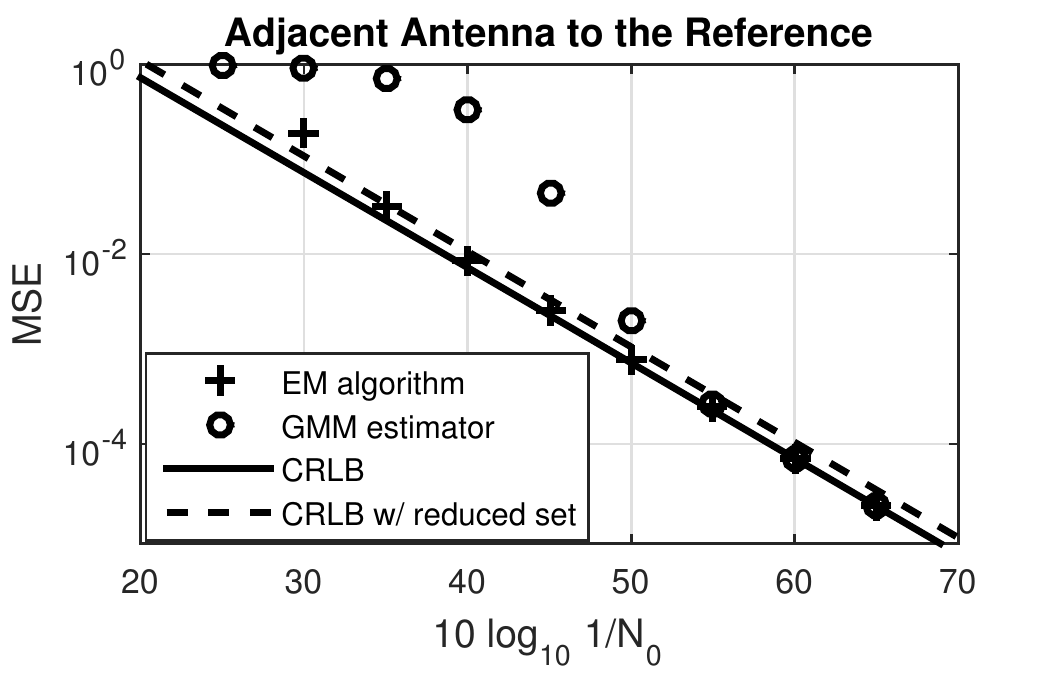}\hspace{0.1cm}
        \includegraphics[scale=.77]{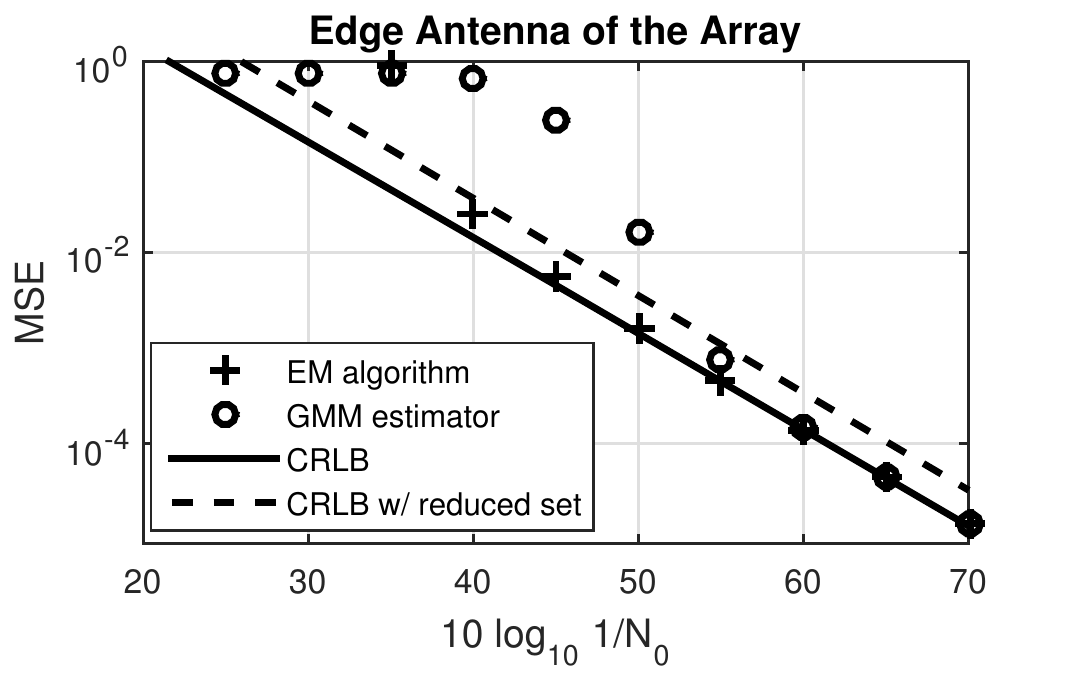}\hspace{0.1cm}
    \end{tabular}
    \caption{MSE of the GMM estimator and the EM algorithm (with $\epsilon=0$), versus their CRLB (solid line), for 2 extreme transceiver cases. Namely, a transceiver associated with an antenna at the edge of the array, and a transceiver associated with an antenna adjacent to the reference. The CRLB plotted by a dashed line is discussed in Sec. \ref{subsec:CalRedMeas}.}
    \label{fig:CRLBvsGMMvsML}
\end{figure*}

Fig. \ref{fig:CRLBvsGMMvsML} compares the MSE  of the estimators with the  CRLB  for two transceiver cases. Both estimators appear to be asymptotically efficient.  Noticeably, the performance gains of the EM algorithm can be grossly superior to the GMM (up to $10$ dB), as it approaches the CRLB at much smaller  values of $N_0$. As mentioned previously, this is mainly because the GMM estimator does not appropriately weight moment conditions with less quality. 

Two remarks about the CRLB itself are now in place. \textit{i)} As mentioned in Appendix B, the assumptions used during the CRLB computations, could result in an underestimated CRLB. 
Indeed, the results in Fig. \ref{fig:CRLBvsGMMvsML} suggest that the assumptions used during the CRLB computations do not affect its final value since the estimators' MSE asymptotically converges to the computed CRLB. This is convenient since (asymptotically) efficient estimators can still be built with limited information.  \textit{ii)} It was assumed that $\phi_{m,n}$ - the phase of $\bar{h}_{m,n}$ - is known during the CRLB computations, although it is originally modeled as a random variable in Sec.\ref{subsec:Modelingcoup}. 
However, if  $\phi_{m,n}$ is assumed to be known, the CRLB is independent of the value of $\phi_{m,n}$. This is because a phase rotation in $\boldsymbol{\mu}_{n,m}$, does not influence (\ref{eq:CRLBgen}), due to the  structure of $\boldsymbol{\Sigma}^{-1}$. Thus, any realization of $h_{m,n}$ - from the model proposed in Sec. \ref{sec:ModMutualCoupling} - provides the same CRLB result.

From the previous two remarks and standard estimation theory \cite{Kay}, it follows that the (narrowband) calibration error - in the high SNR regime - produced by the studied estimators can be well modeled as a multivariate zero-mean Gaussian distribution with covariance matrix given by the transformed inverse Fisher information matrix, found in (\ref{eq:CLRB}). The Gaussianity of the calibration error is further verified (experimentally) in Sec. \ref{subsec:CalErrorMod}.



\subsubsection{Convergence of the EM algorithm}
\label{subsubsec:ConvEM}
\begin{figure*}
    \center
    \begin{tabular}{cc}
        \includegraphics[scale=.75]{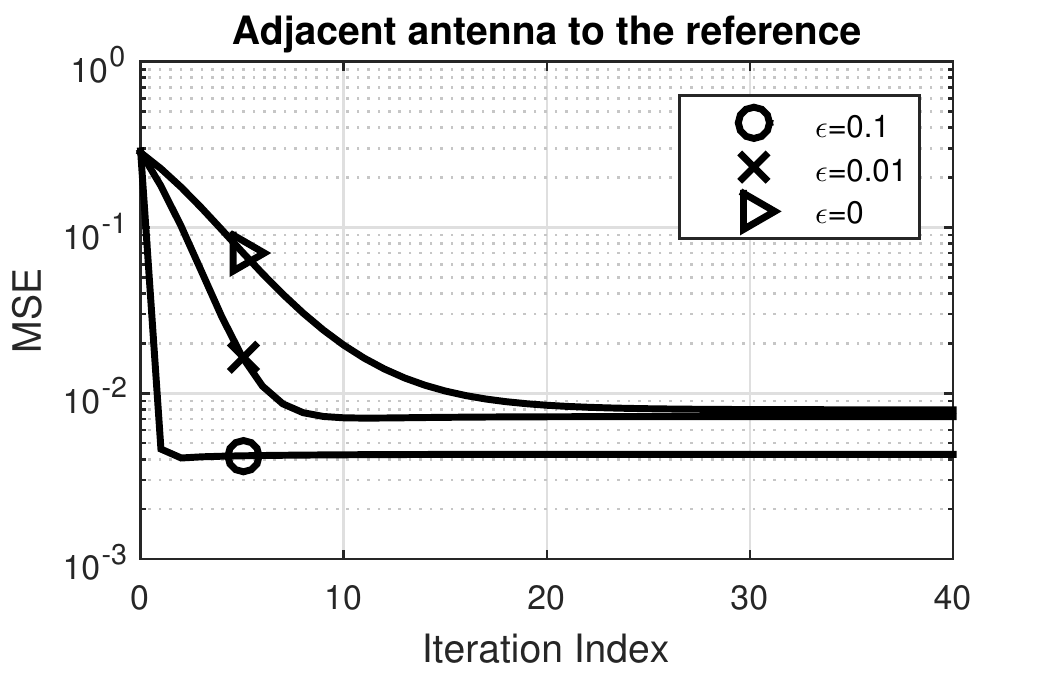}\hspace{0.1cm}
        \includegraphics[scale=.75]{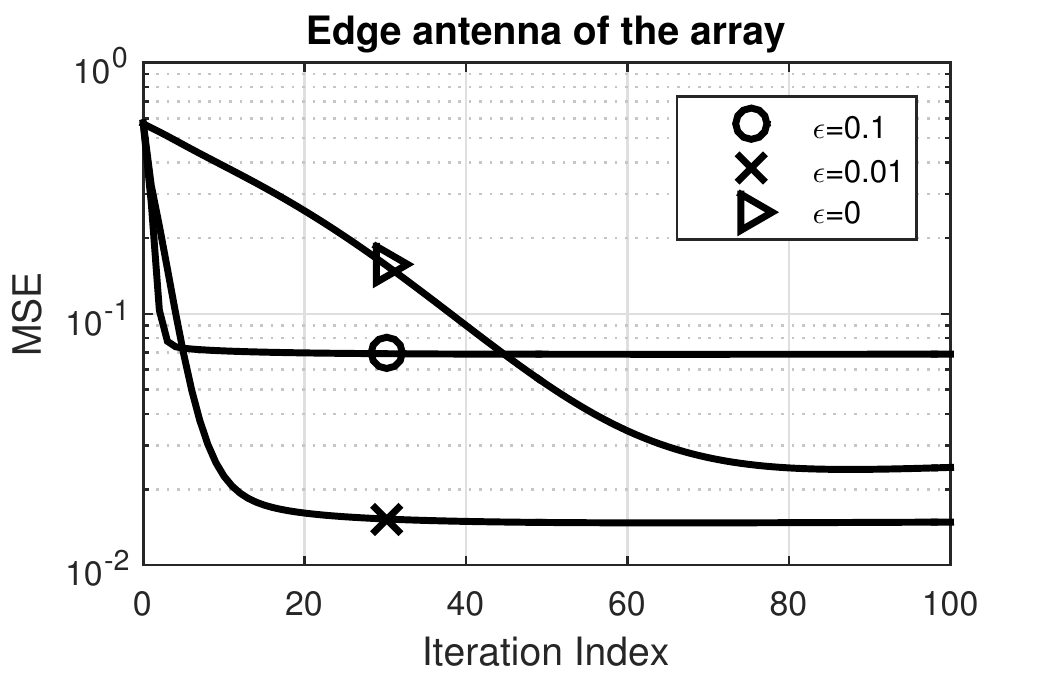}\hspace{0.1cm}
    \end{tabular}
    \caption{ MSE per iteration of the EM algorithm, for different regularization constants $\epsilon$. The plots are for $N_0=-40$ dB, and the remaining simulation settings are the same as Fig. \ref{fig:CRLBvsGMMvsML}. Note the different scales of the plots. }
    \label{fig:NiterationsML}
\end{figure*}

The convergence is analyzed for $N_0=-40$ dB, which from Fig. \ref{fig:CRLBvsGMMvsML} appears to be a region where EM-based estimation provides significant gains compared to GMM. Fig. \ref{fig:NiterationsML} illustrates the role played by the regularization constant $\epsilon$ in terms of convergence rate and MSE. 
Noticeably, the higher $\epsilon$ the faster the algorithm appears to converge. The number of iterations until convergence $N_{\rm ite}$ is seen to be much smaller than $M$ with large enough $\epsilon$ (i.e., around $5$ iterations when $\epsilon=0.1$).\footnote{If, instead, the initial guess is chosen randomly (e.g., calibration coefficients with unit-norm and i.i.d. uniform phases) then our simulations indicate that the order of $N_{\rm ite}$ is $\mathcal{O}(M)$.} 
However, increasing $\epsilon$ indefinitely is not an option as it  degrades the performance. Moreover, the results also indicate that proper tuning of $\epsilon$ can  provide MSE gains compared to the unregularized case which is asymptotically efficient (notice that this does not conflict with the CRLB theorem, as an estimator built with $\epsilon \neq 0$  is not necessarily  unbiased).  This was - to some extent - expected due the benefits of Ridge Regression  as discussed in Sec.\ref{subsec:EMalg}. 

With that, we identify that a fine tuning of $\epsilon$ can provide many-fold improvements. We note that in the literature there is a number of approaches available that deal with optimization of regularization constants in standard (non-iterative) LS problems \cite{Bishop}. However, they are not directly applicable to this work as they typically optimize  single error metrics, and are in general computationally expensive. Here, our main use for $\epsilon$ is to accelerate the convergence and provide estimation robustness to the algorithm, all achieved at no complexity cost. For this matter, we treat $\epsilon$ as a hyperparameter (an approach widely adopted in regularized LS adaptive filtering \cite{Haykin}). Further investigation on fully automatizing the EM algorithm is an interesting matter of future work. 

For the remainder of the paper, we set $\epsilon = 0$ and proceed accordingly, for simplicity.

\subsubsection{Simulation Setup for Sum-rate Capacity Analysis }

The same parameter setting as in Sec. \ref{Sec:SimSetup} is kept in this setup,  and the remaining simulation framework is defined next.  

We assume that the uplink channel ${\bf H}_{\rm UP}$ is perfectly know to the BS, and that there are two noise sources in the system. The first noise source is downlink additive noise modeled by $\bf w'$, see ({\ref{eq:DownLinkCh}}). Here, $\bf w'$  have i.i.d.  zero-mean circularly symmetric complex Gaussian distributed random entries with variance $N_w$ equal to 1. The same model is used for the entries of the downlink channel matrix ${\bf H}_{\rm DL}$. The second noise source is the error during estimation of $\bf c$ (i.e., calibration error). With that,   the precoded signal ${\bf z'}  = {\bf P x}$ is subject to  calibration errors.  The transmit power constraint  $\mathbb{E}{\left\{||{\bf z'}||^2\right\}}= K$ is used. 
Also, we set $K=10$ single antenna users, and assume $t^U_k = t^B_k$ and $r^U_k = r^B_k$ for sake of simplicity. 

The sum-rate capacities \cite{Paulraj} are evaluated for different calibration cases. More specifically, when no calibration is employed (i.e., $\hat{c}_m =1$), when calibration is performed with the GMM or the EM algorithm, for the case of perfect calibration (i.e., $\hat{c}_m = c_m$), and as a baseline, when precoding is performed using the \textit{true} downlink channel ${\bf H}_{\rm DL}$. The analysis is performed with $N_0=-40$ dB, for the reasons mentioned during the convergence analysis.

  \subsubsection{Sum-rate Capacity Results}

\begin{figure*}
    \center
    \begin{tabular}{cc}
        \includegraphics[scale=.74]{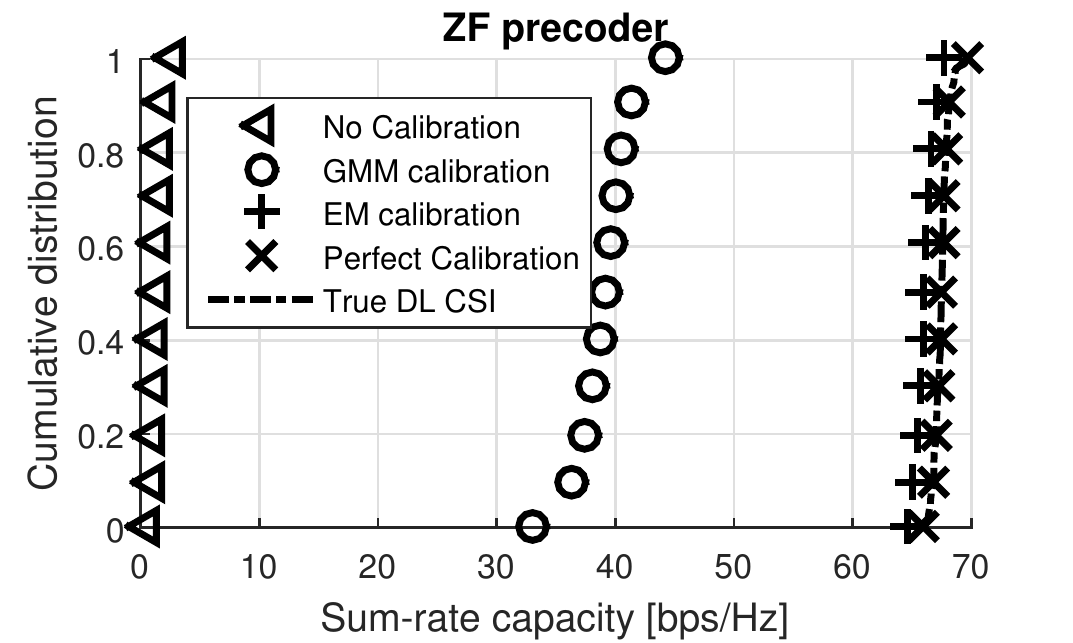}\hspace{0.1cm}
        \includegraphics[scale=.74]{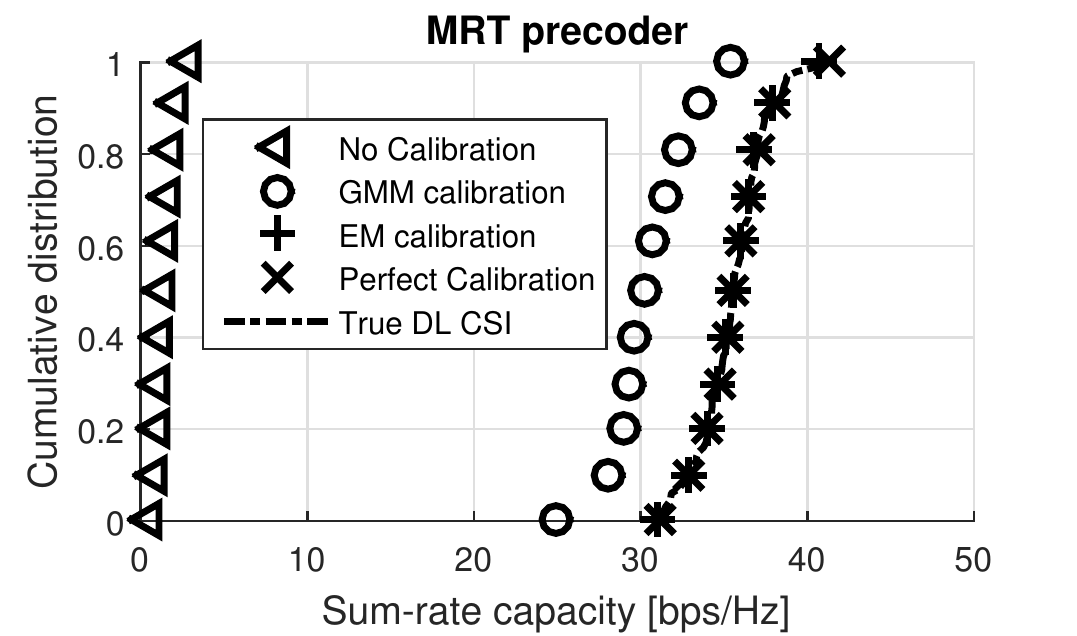}\hspace{0.1cm}
    \end{tabular}
    \caption{ CDFs of the  sum-rates capacities  for different calibration cases. Left) ZF precoder; Right) MRT precoder.
    }
    \label{fig:ZfandMRTcap}
\end{figure*}

Fig. \ref{fig:ZfandMRTcap} shows the obtained sum-rates cumulative distribution functions (CDFs) for different precoding schemes \cite{LUP4305564}. Similarly to the MSE results, EM-based calibration provides significant gains compared to the GMM case. The magnitude of these gains obviously depend on both the calibration (and communication system) setup. For example, there are no sum-rate differences when $N_0 \rightarrow 0$ or $N_0 \rightarrow \infty$, as both GMM and EM approaches converge to that of perfect calibration, or to the uncalibrated case, respectively. Thus, it in only in a certain region of $N_0$ values that EM based calibration provides gains.\footnote{Our analysis based on a wide range of parameter values also indicates  that, in general, stricter calibration requirements need to be met in order to release the full potential of ZF compared to MRT precoding (i.e., no sum-rate difference compared to the perfect calibrated case). Noticeably, this observation is in line with previous calibration studies \cite{7434053}.}

It is interesting that - for this setup - there is no fundamental loss in capacity between this calibration approach (i.e., precoding with perfectly calibrated uplink CSI) and precoding with the true downlink CSI. 
Quantifying this loss is out of scope of this work, however, the interested reader is referred to \cite{7018038} for an overview on the loss of different types of reciprocity calibration. We now finalize the section with two interesting remarks.

\subsection{Remark 1: Calibration with Reduced Measurement Sets }
\label{subsec:CalRedMeas}
There are several benefits of using a reduced measurement set for calibration (e.g., by only relying on high quality measurements). This is possible as long as (\ref{eq:EqModel}) is not under-determined. As an illustrative example, the dashed line in Fig. \ref{fig:CRLBvsGMMvsML} shows the CRLB when a reduced measurements set - comprising the measurements between antenna pairs whose elements are distanced by at most $1/\sqrt{2}$ wavelengths - is used. The number of measurement signals in this case drops from $M(M-1)$ to less than $8M$, since one antenna signals to, at most, $8$ other antennas. The  performance loss turns out to be insignificant, i.e. $2$ dB for the neighbor case and $4$ dB for the edge case, considering the number of signals discarded.  This indicates that the channels between neighbor antennas, which are dominated by mutual coupling, are the most important for calibration. Thus, there is an interesting  trade-off between the asymptotic performance of an estimator and its computational complexity (proportional to the number of measurements).

Another benefit of using reduced measurement sets is a possible reduction of resource overhead dedicated for calibration. This can be very important from a system deployment point-of-view. To finalize, we remark that ML closed form estimators can be also reached when reduced measurement sets are used. This can be the case for the current (general) calibration setup when a reduced set of measurements is used, or for the case of working with a full set of measurements when  the calibration setup is a special case. An example of the latter is given next.

\subsection{Remark 2: Closed-form Unpenalized ML Calibration for Linear arrays }
\label{sec:LinearArray}
Consider an $M$-antenna linear array, and let $m$ index the antennas in ascending order starting at one edge of the linear array. Assume that mutual coupling only exists between adjacent antenna elements, and that the channel between any other antenna pairs is weak enough so that it can be neglected without any noticeable impact on performance. We summarize our findings in Proposition 1.

\textit{Proposition 1:}  Using a reference antenna as a starting point, say $c_{1}=1$, the unpenalized ML solution for any $c_{\ell+1}$, with $1\leq \ell \leq M-1$, can be obtained sequentially by
\begin{equation}
\label{eq:closedFormML}
\hat{c}_{\ell+1} = \hat{c}_{\ell}\frac{ y^*_{\ell+1,\ell} y_{\ell,\ell+1}   }{ |y_{\ell+1,\ell}|^2 }.
\end{equation}

\iftoggle{2Collumn}{
    \textit{Proof:} See Appendix C. \hspace*{45mm} \rule{2.5mm}{2.5mm}}{
\textit{Proof:} See Appendix C. \hspace*{116mm} \rule{2.5mm}{2.5mm}}

We can also deduce the following interesting corollary.

\textit{Corollary 1:} For any of the two constraints considered in (\ref{eq:GMMeCost}), the GMM (vector) estimator  coincides with (\ref{eq:closedFormML}) up to a common complex scalar.

\iftoggle{2Collumn}{
    \textit{Proof:} See Appendix C. \hspace*{45mm} \rule{2.5mm}{2.5mm}}{
    \textit{Proof:} See Appendix C. \hspace*{116mm} \rule{2.5mm}{2.5mm}}

\section{Validation of the calibration method in a massive MIMO testbed} 
\label{sec:Validation}
In this section, we detail the experiment performed to validate the proposed mutual coupling based calibration method. More specifically,  we implemented it in a software-defined radio testbed, and performed a TDD transmission from 100 BS antennas to 3 single antenna terminals. 

Note that the analysis conducted in this section and in Sec. \ref{sec:WB_EM} is  measurement based. As stationarity is assumed in the analysis, we monitored the system temperature throughout the measurements and verified no significant changes. We also made  an effort to keep static  propagation conditions, and performed  the experiments at late hours in our lab with no people around.
\subsection{Brief Description of the Testbed}

Here we briefly outline the relevant features of the testbed for this work. Further information can be found in \cite{BasestationPaper}.

\subsubsection{Antenna/Transceiver setup}
\label{sub:ANteTrans}

The BS operates with 100  antennas, each antenna connected to one distinct transceiver.
For simplicity, the same transceiver settings (e.g.,  power amplifier gain and  automatic gain control) are used in  both calibration and data communication stages for all radio units. This ensures that the analog front-ends yield the same response during both stages,  thus the estimated calibration coefficients are valid during the communication stage.


\subsubsection{Synchronization of the radios}

Time and Frequency synchronization is achieved by distributing reference signals to all radio units.  However, this does not guarantee phase alignment between all BS transceiver radio chains which motivates reciprocity calibration. 


\subsection{Communication  Protocol used}


\footnotesize

\begin{table}[t]
    \caption{High-level OFDM parameters}
    \noindent\begin{tabular*}{\columnwidth}{@{\extracolsep{\stretch{1}}}*{7}{l}@{}}
        \toprule
        \textbf{Parameter} & \textbf{Variable} & \textbf{Value} \\
        \midrule
        Carrier frequency &  $f_{\rm c}$ & 3.7\;GHz \\
        Sampling Rate & $F_{\rm s}$ & 7.68\,MS/s \\
        FFT Size & $N_{\rm FFT}$ & 2048 \\
        \# Used sub-carriers & $N_{\rm SUB}$  & 1200 \\
        \bottomrule
    \end{tabular*}
    \label{table:SystemParam}
\end{table}

\normalsize


Once the measurements to construct the observation matrix $\bf Y $ are performed, $\bf c$ is estimated using the unpenalized EM algorithm.  The following sequence of events is then performed periodically:
\subsubsection{Uplink Channel  Estimation and Calibration}Users simultaneously transmit frequency orthogonal pilot symbols. The BS performs LS-based channel estimation, and interpolates the estimates between pilot symbols. Reciprocity calibration is then performed independently per subcarrier, i.e. as in (\ref{eq:lastEq}), for coherence purposes with Sec. \ref{sec:ChannelModel}. This calibrated version of the downlink channel is then used to construct a ZF precoder.

\subsubsection{Downlink channel estimation and data transmission} Downlink pilot symbols are precoded in the downlink  and each user performs LS-based channel estimation. Using the estimates, each user recovers the payload data using a one-tap equalizer.


We note that 4-QAM signaling per OFDM sub-carrier is used for  uplink channel estimation and data transmission. The main parameters are shown in Table \ref{table:SystemParam}. Further information on the signaling protocol (e.g.,  uplink/downlink frame structure or uplink pilot design) is found on \cite{BasestationPaper}. 

\subsection{Measurement Description}

The setup used in our experiments is shown in Figure \ref{fig:DownlinkSetup}. Although not being a typical propagation scenario found in  cellular systems,   this extreme setup - closely located users under strong line-of-sight conditions - requires  high calibration requirements to be met if spatial separation of users is to be achieved. In addition, we use ZF precoding  as it is known to be very sensitive to calibration errors \cite{7018038}. 

The EVM \cite{schenk2008rf} of the downlink equalized received samples at  each mobile station was evaluated, and used as performance metric for validation purposes. The rationale is that, with multiple mobile terminals, calibration errors are translated into downlink inter-user interference (and loss of array gain), which increases the EVM. Letting $r$  be the downlink equalized received sample when symbol $s$ is transmitted, the EVM is defined as
\begin{equation}
\label{eq:EVMs}
\mathrm{EVM} = \mathrm{E}\left\{ \frac{ |r-s|^2 }{ |s|^2}\right\},
\end{equation}
where the expectation is taken over all system noise sources (e.g., hardware impairments and thermal noise). Our estimate of (\ref{eq:EVMs}) was obtained by averaging realizations of $|r-s|^2 / |s|^2$ over all OFDM sub-carriers and over received OFDM symbols.

We estimated the EVM for different energy values of the uplink pilots and calibration signals. We do so in order to be able to extract insightful remarks for the analysis of the results. In particular, letting $E_{\rm Pilot}=\mathrm{E} \left\{ p_k p_k^* \right\}$  in (\ref{eq:UpLinkCh}) denote the energy of the uplink pilot, which, for simplicity, is the same for all users, and let $E_{\rm Cal}$ denote the energy of the sounding signal $s_{m}$ in (\ref{eq:SystemModel33}), we estimated the EVM for a 2-dimensional grid of $E_{\rm Pilot}$ and $E_{\rm Cal}$. The results reported next are given with respect to the relative energies $Er_{\rm Pilot}= E_{\rm Pilot} / E^{\rm max}_{\rm Pilot} $ and $Er_{\rm Cal}= E_{\rm Cal} / E^{\rm max}_{\rm Cal}$, where $E^{\rm max}_{\rm Pilot}$ and $ E^{\rm max}_{\rm Cal}$ are the maximum energies of the uplink pilot and calibration signal used in the experiments. Other systems parameters (e.g., transmit power in the downlink)  were empirically set and kept constant throughout the experiment.

\subsection{Validation Results}

\begin{figure*}[t]
    \center
    \includegraphics[scale=.80]{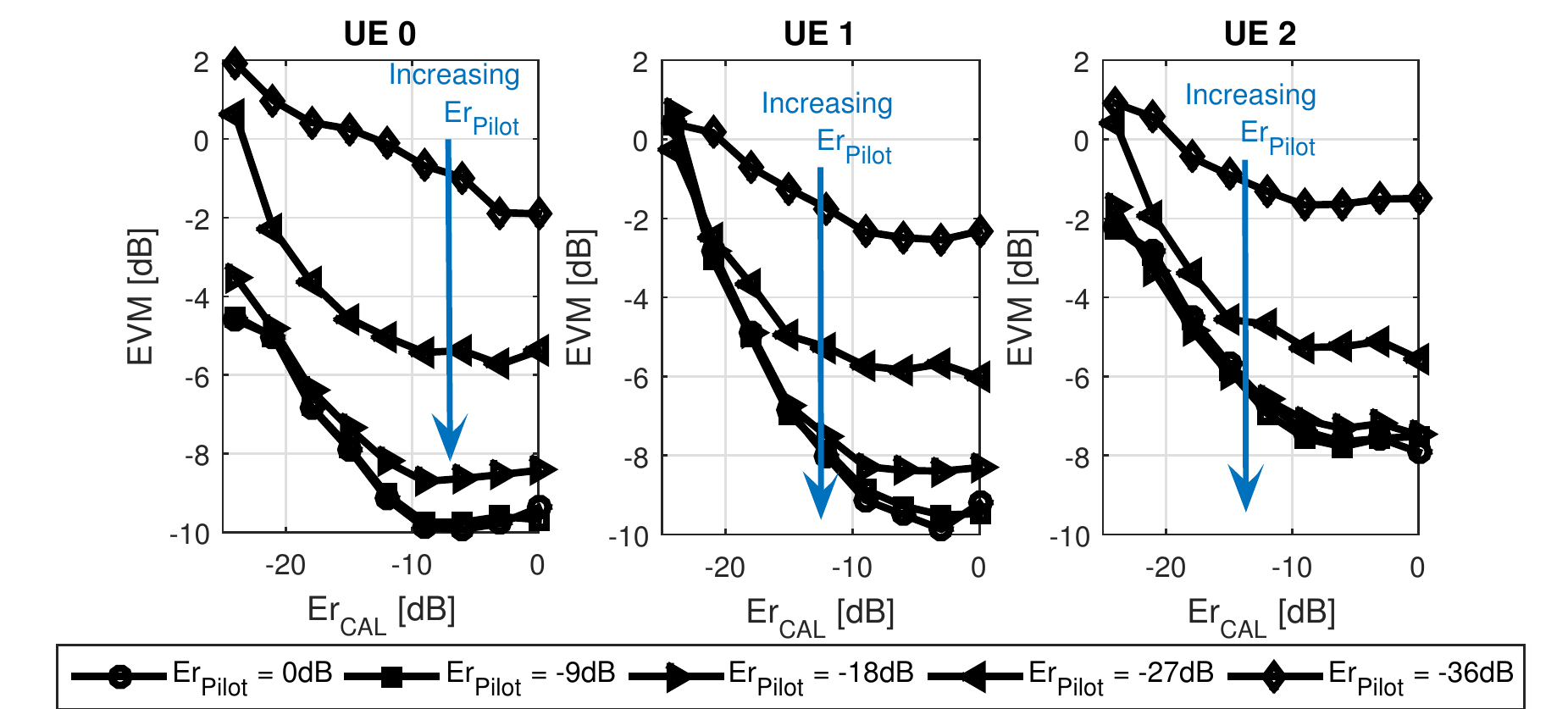}
    \caption{Measured EVM at each of the three user terminals during a massive MIMO downlink transmission.}
    \label{fig:AntennaCouplingNo2}
\end{figure*}

Fig.  \ref{fig:AntennaCouplingNo2} shows the measured EVMs for the $3$ user terminals in our experiment. 
Before discussing the results, we remark that analyzing the EVM when $Er_{\rm Cal}$ is reduced beyond $-30$ dB is not of fundamental interest, as it approaches the  uncalibrated case (where high EVMs are to be expected). Overall, a positive trend is observed with increasing  $Er_{\rm Cal}$ until $-10$ dB. This reflects the BS ability of spatially separating users which increases with increasing the calibration quality. The fact that downlink EVMs down to $-10$ dB are achieved, which are much smaller than the EVMs when $Er_{\rm Cal}= -30$ dB, i.e. close to the uncalibrated case, motivates our validation claim.

It is possible to observe a saturation of the EVMs at high enough $Er_{\rm Cal}$ and $Er_{\rm Pilot}$ for all user cases. This is an expected effect in practical systems. Explained briefly, system impairments other than the calibration or the uplink channel estimation error, become the dominant error sources that bound the EVM performance\footnote{Mobile terminals error sources (e.g.,  in-phase and quadrature imbalance or thermal  noise) qualify for such impairments. For a given downlink transmit power, it is straightforward to understand how such impairments bound the downlink EVMs regardless of the calibration and uplink estimation quality.}. Remarkably, this saturation effect implies that the calibration SNR - available in a practical array as ours - is sufficiently large not to be the main impairment to constrain the system performance. Mutual coupling channels are thus reliable (and reciprocal enough), so that they can be used for signaling in order to calibrate the system.\footnote{We note there exists an interesting theoretical trade-off between the calibration quality and the capacity of downlink channels with respect to the strength of mutual coupling. In practice, the proposed calibration method can be used in compact antenna arrays with very low coupling (say $-30$ dB between adjacent elements) provided that the transmit power during calibration is sufficient to provide good enough estimation SNR. In such a setup, the impact of coupling in the capacity is negligible.}

\section{Aspects of Wideband Calibration and Error Modeling}
\label{sec:WB_EM}

A short summary of this section follows. Using the measurements from the Sec. \ref{sec:Validation}, we treat the estimated calibration coefficients across OFDM sub-carriers as realizations of a discrete stochastic process.  Using low rank approximation theory, we propose a parametrized low dimensional basis that  characterizes the subspace spanned by this process accurately. Based on the reduced basis, we propose a wideband estimator that averages out the calibration error across frequency. Using the wideband estimator results, we validate the narrowband calibration error model proposed in Sec. \ref{subsubsec:MSEvsCRLB}. We remark that our experiment makes use of a  bandwidth of $ F_s N_{\rm sub} / N_{\rm FFT} = 4.5$MHz.

\subsection{Wideband Remarks for the Calibration Coefficients}

Denote the calibration coefficient of BS antenna $m$ at the $k$th OFDM sub-carrier as $C_m[k] = t_m^k / r_m^k $. The variable $\hat{C}_m [k]$ is the estimate of $C_m[k]$ at sub-carrier $k$ - obtained, e.g., with the EM algorithm - and is modeled as
\begin{align}
\label{eq:CCsMod}
    \hat{C}_m [k] = & C_m[k] + E_m[k]  \nonumber \\
								  =& |C_m[k]|\exp(j2\pi\zeta_m[k]) + E_m[k]
\end{align}
where  $E_m[k]$ is an i.i.d. random process representing the calibration error which is assumed zero-mean and independent of $C_m[k]$. 
Let the random phasor process $\exp(j2\pi\zeta_m[k])$ in ($\ref{eq:CCsMod}$) absorb the phase shift stemming from the arbitrary time that a local oscillator needs to lock to a reference signal. Such phase shift is often modeled as uniformly distributed, and thus
\begin{equation}
\label{eq:ZR-mean}
\mathrm{E}\left\{\exp(j2\pi\zeta_m[k])\right\}=  0.
\end{equation}
Moreover,  since local oscillators associated with different transceivers  lock at arbitrary times, it is safe to assume 
\begin{equation}
\label{eq:UncProc}
\mathrm{E}\left\{\exp(j2\pi\zeta_m[k_1])   \exp(-j 2\pi\zeta_n[k_2])  \right\}=  0, \; m\neq n.
\end{equation}
Not making further assumptions on the statistics of $\hat{C}_m [k]$, we now proceed with a series expansion, but before doing so we make one last remark. The series expansion conducted next is performed based on measurements from the 100 testbed transceivers, and serves as an example approach to obtain a suitable  basis for $\hat{C}_m [k]$. This can well apply to mass-production transceiver manufactures that can reliably estimate the statistical properties of the hardware produced. However, as our testbed operates with relatively high-end transceivers - compared to the ones expected to integrate commercial massive MIMO BSs - the dimensionality of the subspace verified in our analysis might be underestimated. Intuitively, the higher transceiver quality, the less basis functions are needed to accurately describe $\hat{C}_m [k]$. Nevertheless, the upcoming remarks apply for  smaller bandwidths - than $4.5$MHz - depending on the properties of the transceivers.

\subsection{Principal Component Analysis}

From the assumption (\ref{eq:ZR-mean}), it follows that the element at the $v_1$th row and $v_2$th column of the covariance matrix ${\bf K}_m$ of $\hat{C}_m [k]$ is defined as 
\begin{equation}
\label{eq:Kmat}
[{\bf K}_m]_{[v_1,v_2]}= \mathrm{E}\left\{ \hat{C}_m[v_1] \; \hat{C}^*_m[v_2] \right\}.
\end{equation}
From the assumption (\ref{eq:UncProc}), it follows that the principal components of $\hat{C}_m[k]$ are obtained by singular value decomposition (SVD) of ${\bf K}_m$ only \cite{van2004detection}. Let the SVD of ${\bf K}_m$ be written as
\begin{equation}
\label{eq:SVD}
{\bf K}_m = \sum_{i=1}^{\rm N_{SUB}} {\bf u}^{ m}_i \lambda^{ m}_i ({\bf u}^{ m}_i)^{ H},
\end{equation}
where  $\left\{  {\bf u}^m_i \right\}_{i=1}^{\rm N_{SUB}}$ are the principal components, and $  { \lambda}^m_i $ is the power (variance) of the coefficient obtained from projecting $\hat{C}_m [k]$ into  ${\bf u}^m_i$. We  use the convention $ \lambda^m_1 \geq  \lambda^m_2 \dots \geq \lambda^m_{N_{\rm SUB}}$, and  $ {\bf u}^m_i = \big[ [{u}^m_i[1], \cdots, {u}^m_i[N_{\rm SUB}] \big]^{\rm T} $.
\begin{figure*}[t]
    \centering
    \includegraphics[scale=.83]{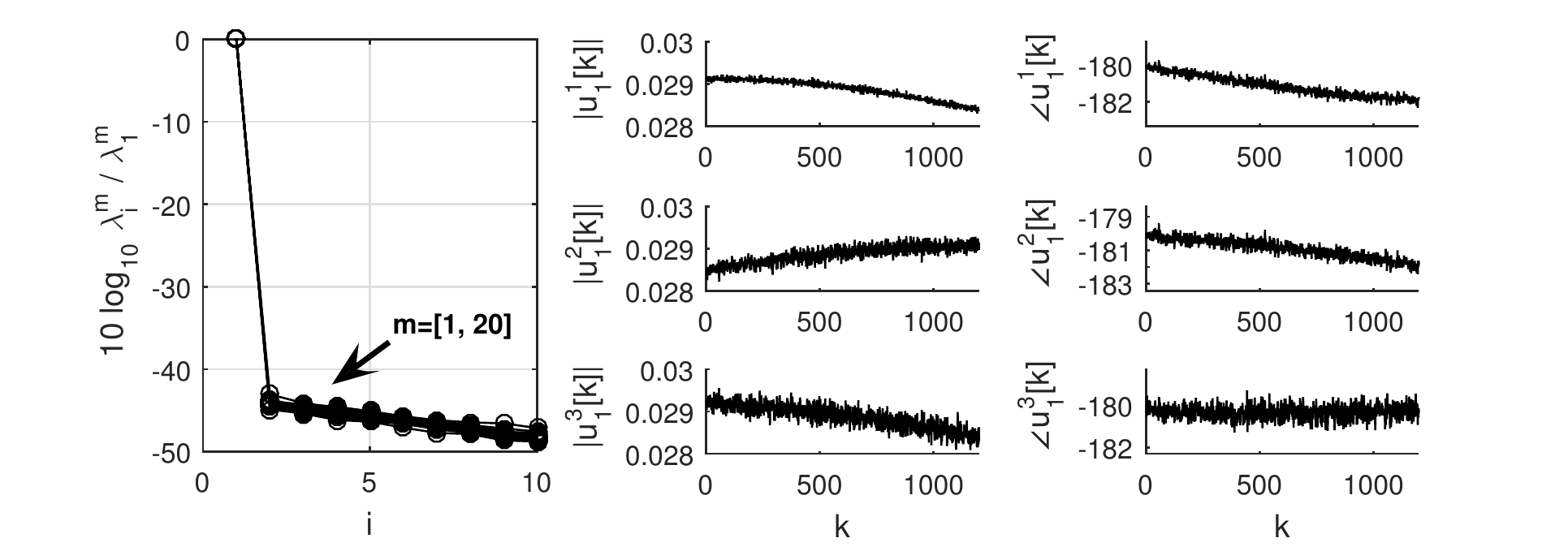}
    \caption{ Principal component and coefficients of  $\hat{C}_m [k]$. \textit{Left)} The 10 strongest normalized singular values for 20 transceivers; \textit{Middle)} Magnitude of the principal component for 3  transceivers; \textit{Right)} Phase of the principal component for 3  transceivers.} 
    \label{fig:SVD}
\end{figure*}
Fig. \ref{fig:SVD} shows several coefficients and basis functions of the expansion, that were estimated based on 100  realizations of $\hat{C}_m [k]$, each measured with $Er_{\rm Cal} = 5$ dB (which from Fig. \ref{fig:AntennaCouplingNo2}  provides a relatively high calibration SNR). Noticeably, it appears that all processes (one per transceiver) live mostly in a one-dimensional sub-space and thus can be well described by their first principal component ${\bf u}^{ m}_1$. This fact also indicates that the contribution of the calibration error in the expansion is small, and thus the first principal component  of $\hat{C}_m [k]$ is also representative for the true coefficients $C_m [k]$.

Visual inspection indicates that both magnitude and phase of the first principal component can be well approximated with a linear slope across frequency. The inherent error of this approximation is very small compared to the magnitude of the process itself. We note that  this linear trend holds for any transceiver of the array (not only for the ones shown in Fig. \ref{fig:SVD}).


%
\subsection{Wideband Modeling and Estimation}

The previous analysis indicates that any first principal component can be well described by a linear magnitude slope  $\gamma_m$, and a linear phase  $\xi_m$ across frequency. Such properties are well captured by the  Laplace kernel 
$\exp( (\gamma_m + j2\pi \xi_m)k ),$
for small values of $|\gamma_m|$ (since the range of $k$ is finite). The final parameter to model a  realization of  the process is the complex offset $A_m$. With that, the general model  (\ref{eq:CCsMod}) can thus be re-written as
\begin{equation}
\label{eq:ModelC}
 \hat{C}_m[k]  =    A_m\exp( (\gamma_m + j2\pi \xi_m)k ) + w_m[k], 
\end{equation}
where $w_m[k]$ is a random process that  absorbs: the calibration error $E_m[k]$,  the error due to the low rank approximation, and the error due to the linear modeling of the first principal component ${\bf u}^{ m}_1$.  Given an observation $\{ \hat{C}_m[k]\}_{k=1}^{ N_{\rm SUB}}$, the ML estimator of $A_m$, $\xi_m$ and $\gamma_m$, namely, $\hat{A}_m$, $\hat{\xi}_m$ and $\hat{\gamma}_m$ is straightforward to derive \cite{Kay}. Thus, we define the wideband  estimator of $\hat{C}_m[k]$ as
\begin{equation}
\label{eq:MLfit}
\hat{C}_m[k]^{ \rm WB} = \hat{A}_m\exp( (\hat{\gamma}_m + j2\pi \hat{\xi}_m)k ).
\end{equation}

\begin{figure}[t]
    \centering
    \includegraphics[scale=.80]{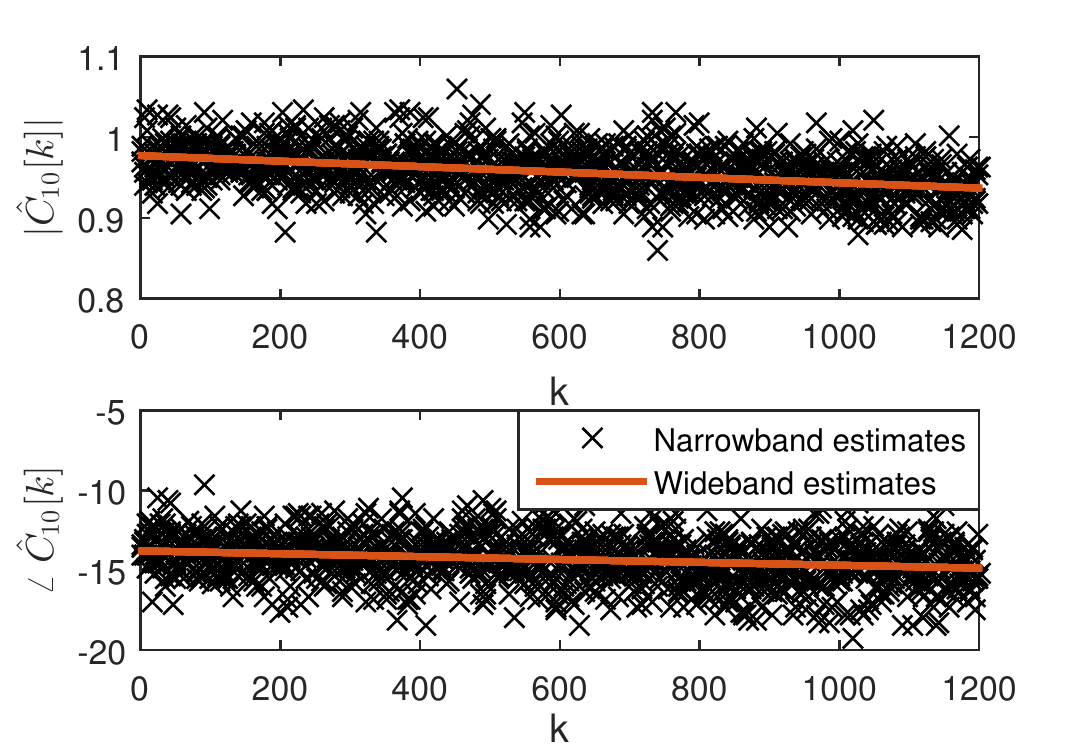}
    \caption{ A  realization of the narrow-band estimator $\hat{C}_m[k]$, and the proposed wideband estimator $ \hat{C}_i[k]^{\rm WB}$.}
    \label{fig:MLfit}
\end{figure}

For illustration purposes, a  realization of the ML wideband estimator $\hat{C}_m[k]^{\rm WB}$ is contrasted with that of the narrow-band estimator $\hat{C}_m[k]$ in Fig. \ref{fig:MLfit}. The obtained error reduction is evident.

\subsection{A Model for the Calibration Error}

\label{subsec:CalErrorMod}
Here, we use the wideband estimator results to verify the Gaussianity of the narrow-band calibration error proposed in Sec. \ref{subsubsec:MSEvsCRLB}.  This is done under the two following main assumptions.

\textit{1) The  residual process $ E_m[k] =  \hat{C}_m[k] - C_m[k]$ is well described by  $ \hat{E}_m[k] =  \hat{C}_m[k] -  \hat{C}_m[k]^{\rm WB}$}. This is reasonable if $ \mathrm{E}\left\{ | \hat{C}_m[k]^{\rm WB} - C_m[k] |^2 \right\} \ll \mathrm{E}\left\{ | \hat{C}_m[k] - C_m[k] |^2 \right\} $. To justify, the estimation gains scale linearly in the number of realizations \cite{Kay}, which is $ N_{\rm SUB}=1200$ in this case. Assuming that: the estimation error is  independent across realizations,  the underlying model (\ref{eq:ModelC}) describes the first principal component well, and the low rank approximation error is minuscule, there are  gains  of $10\log_{10}N_{\rm SUB}\approx 30$ dB which justify the first main assumption.

\textit{2) The residual process $ E_m[k] $ is ergodic}.\footnote{Ergodicity is necessary since each (independent) measurement of $\hat{C}_m[k]$ takes about ten minutes with our test system (due to the locking time of the local oscillator to the reference signal). As potential system temperature drifts during the measurements can result in varying statistical properties, it is safer to perform the  analysis based on one solely realization of $E_m[k]$.} This is met if $E_m[k]$ is stationary and the ensemble of $N_{\rm SUB}$ samples is representative for statistical modeling. The former holds for  small OFDM bandwidths (e.g., $4.5$ MHz) as  the hardware impairments do not vary significantly across the band. The latter is also met, as we have $N_{\rm SUB}=1200$ narrow-band estimators whose estimated errors  $\{ \hat{E}_m[k] \}_{k=1}^{N_{\rm SUB}}$ were found to be mutually uncorrelated.

\begin{figure}[t]
    \centering
    \includegraphics[scale=.79]{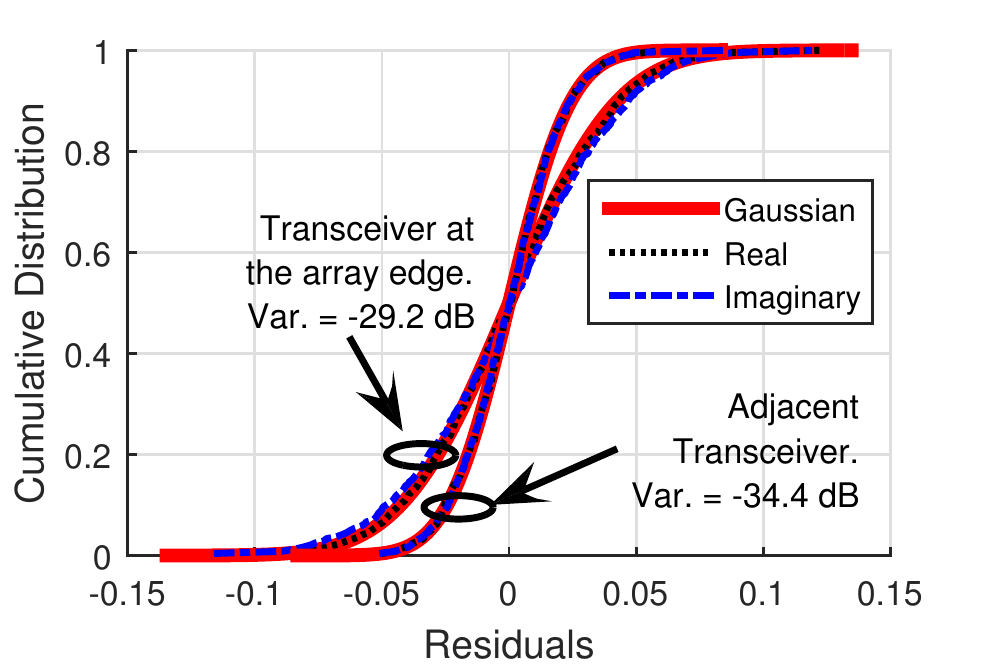}
    \caption{ Empirical CDFs for the real and imaginary parts of the  calibration error,  for a transceiver at the edge of the array, and for an adjacent transceiver to the reference antenna. A Gaussian CDF of equal variance is plotted for both cases for comparison.}
    \label{fig:CDFerror}
\end{figure}

Fig. \ref{fig:CDFerror} shows the empirical CDF of both real and imaginary parts of $\{ \hat{E}_m[k] \}_{k=1}^{N_{\rm SUB}}$ - which we found to the uncorrelated - for two transceiver cases.
Each of the empirical CDFs is contrasted with a zero-mean Gaussian distribution of equal variance. Overall, the empirical CDFs for both transceivers resemble a Gaussian CDF extremely well. The Gaussianity of the calibration error was further verified by passing a Kolmogorov-Smirnov  test with $0.05$ significance level \cite{daniel1990applied}.  We note that these observations hold not only for the two transceivers in Fig. \ref{fig:CDFerror}, but for all transceivers of the array.  Noticeably, the empirical distribution of the calibration error is in line with the  asymptotic properties of ML estimators, i.e. the error can be modeled by an additive  zero-mean Gaussian multivariate.  The final element for a full  characterization  is its covariance matrix, relating the errors across antennas. A good approximation (at high SNR) is the inverse of the transformed Fisher Information matrix in (\ref{eq:CLRB}). Noticeably, future calibration works can benefit from the convenience of safely assuming a non-white Gaussian calibration error. 

\section{Conclusions} \label{sec:conclusions}

We have proposed and validated a convenient calibration method which rely on mutual coupling  to enable the reciprocity assumption in TDD massive MIMO systems. We verified that in a practical antenna array, the channels due to mutual coupling are reliable and reciprocal enough, so that they can  be used for signaling in order to calibrate the array. 

The iterative ML algorithm  is asymptotically efficient and outperforms current state-of-the-art estimators in an MSE and sum-rate capacity sense. Further improvements - in terms of MSE and convergence rate - can be harvested by proper tuning of its regularization hyperparameter.

The calibration error can be further reduced by proper averaging over the radio bandwidth. More importantly, it did not stand as the main impairment to constraint the performance of the system, from our experiments. Our measurements also verified that the narrow-band calibration error (at high SNR) is Gaussian distributed, which is coherent with the theory of the estimator proposed. The convenience of safely assuming  a non-white Gaussian calibration error can, hopefully, open the door for future analytical studies of calibrated TDD massive MIMO systems.


\section*{Acknowledgments}

This work was funded by the Swedish foundation for strategic research SSF, VR, the strategic research area ELLIIT, and the E.U. Seventh Framework Programme (FP7/2007-2013) under grant agreement n 619086 (MAMMOET). We also thank the Comm. Systems group in Bristol University, for letting us replicate several of our results in their testbed.

\section*{Appendix A: Equivalent Channel Matrices}
\label{Appendix:A}
Here we show the structure of the equivalent models. Define the column vector  ${\bf\Psi}_{m} = \left[ \psi_{1,m}   \dots \psi_{m-1,m} \; \psi_{m+1,m} \dots \psi_{M,m} \right]^T$. The equivalent channel matrix in (\ref{eq:EqModel10}) is written as
\begin{equation}
\label{eq:PsiEqStruct}
{\bf\Psi_{\rm eq}({\bf \tilde{\Psi}})}  =
{\rm diag}\left\{ {\bf\Psi}_{1}, {\bf\Psi}_{2}, \hdots, {\bf\Psi}_{M}  \right\}.  
\end{equation}
Now define
\begin{equation}
\label{eq:smallC}
{\bf \boldsymbol{\bar{c}}}_{n,m} = [ c_{n} \; c_{m} ]^T. 
\end{equation}
 Noting that  $\psi_{m,n} = \psi_{n,m}$, the equivalent  matrix and the parameter vector in (\ref{eq:EqModel11}) are written as

\begin{equation}
\label{eq:CeqStruct}
{\bf C_{\rm eq}({\bf c })}  = {\rm diag}\left\{ {\bf \boldsymbol{\bar{c}}}_{1,2},\cdots, {\bf \boldsymbol{\bar{c}}}_{1,M} , {\bf \boldsymbol{\bar{c}}}_{2,3}, \cdots , {\bf \boldsymbol{\bar{c}}}_{2,M} , \cdots \right\},
\end{equation}
and
\begin{equation}
{\bf \tilde{\Psi}}  = \begin{bmatrix}
\psi_{2,1} \dots \psi_{M ,1} \; \psi_{3,2} \dots \psi_{M ,2} \dots \psi_{M ,M-1}
\end{bmatrix}^T.
\end{equation}

\section*{Appendix B: The Cram\'{e}r-Rao Lower Bound}
%
Here we compute the  CRLB for  the calibration coefficients $\left\{c_m\right\}\setminus c_\textit{ref}$. The exclusion of  $ c_\textit{ref}$ is justified in the end of the calculations. This is achieved by assuming $t_{ ref}=r_\textit{ref}=1$, and treating $c_\textit{ref} = t_\textit{ref}/r_\textit{ref}$  as known for estimation purposes. Define the $(4M-4) \times 1$ vector
\begin{align}  
\label{eq:ThetaDef}
{\bf \boldsymbol{\theta}}\!=\!\left[ \operatorname{Re}\!\left\{ t_{1} \right\} \; \operatorname{Im}\!\left\{ t_{1} \right\} \; \operatorname{Re}\!\left\{ r_{1} \right\} \;\operatorname{Im}\!\left\{ r_{1} \right\} \; \operatorname{Re}\!\left\{ t_{2} \right\} \dots  \operatorname{Im}\!\left\{ r_{M} \right\} \right]^T,
\end{align}
where $t_\textit{ref}$ and $r_\textit{ref}$ do not enter. The CRLB for $\left\{c_m\right\}\setminus c_\textit{ref}$ is given by the diagonal entries of the transformed inverse Fisher information matrix \cite{Kay}
\begin{equation}
\label{eq:CLRB}
\text{var} ( \hat{c}_m ) \geq \left[  \frac{ q( \boldsymbol{\theta}) }{\partial \boldsymbol{\theta}}  \textbf{I}^{-1}( \boldsymbol{\theta} ) \frac{ q( \boldsymbol{\theta} )}{\partial \boldsymbol{\theta}}^{\rm H} \right]_{m,m},  \; m\neq  \textit{ref},
\end{equation}
where $\textbf{I}( \boldsymbol{\theta} )$ is the Fisher information matrix of  $\boldsymbol{\theta}$. The transformation of $\boldsymbol{\theta}$ into the calibration coefficients is given by $$ q(\boldsymbol{\theta})  = \left[ \frac{  \operatorname{Re}\!\left\{ t_{1} \right\} + j\operatorname{Im}\!\left\{ t_{1} \right\}  }{ \operatorname{Re}\!\left\{ r_{1} \right\} + j \operatorname{Im}\!\left\{ r_{1} \right\}  } \dots \frac{  \operatorname{Re}\!\left\{ t_{M} \right\} + j \operatorname{Im}\!\left\{ t_{M} \right\}  }{  \operatorname{Re}\!\left\{ r_{M} \right\} + j \operatorname{Im}\!\left\{ r_{M} \right\}  } \right]^T.$$
We now compute  $\textbf{I}( \boldsymbol{\theta} )$. Assuming that $\bar{h}_{m,n}$, $\sigma^2$ and $N_0$ are at hand,\footnote{These assumptions are only used for the CRLB calculations, and were not used to derive any of the estimators. A possible implication is that the CRLB can be underestimated, but we will see that this is not  the case from the simulations' results.} the mean $\boldsymbol{\mu}_{n,m}$ and the covariance matrix $\boldsymbol{\Sigma}_{n,m}$ of ${\bf y}_{n,m} = [y_{n,m} \; y_{m,n}]^T$ are given by
\begin{equation}
\label{eq:subMean}
\boldsymbol{\mu}_{n,m} = \mathrm{E}\left\{ {\bf \boldsymbol{y}}_{n,m} \right\} = \bar{h}_{n,m} \left[ r_nt_m \;  r_mt_n \right]^T,
\end{equation}
\begin{align}
\label{eq:subCov}
\boldsymbol{\Sigma}_{n,m} & = \mathrm{E}\left\{  ({\bf\boldsymbol{y}}_{n,m} - \boldsymbol{\mu}_{n,m}) ({\bf\boldsymbol{y}}_{n,m}-\boldsymbol{\mu}_{n,m})^H \right\} \nonumber \\ 
& =  \begin{bmatrix}
|r_n|^2 |t_m|^2 \sigma^2 + N_0 & r_n t_m r_m^* t_n^* \sigma^2  \\
r_m t_n r_n^* t_m^* \sigma^2 & |r_m|^2 |t_n|^2 \sigma^2 + N_0 \\
\end{bmatrix}.
\end{align}
We can observe that the PDF of ${\bf Y}''$, where $$
{\bf Y}'' = \left[ {\bf\boldsymbol{y}}^T_{1,2} \dots {\bf\boldsymbol{y}}^T_{1,M} \; {\bf\boldsymbol{y}}^T_{2,3} \dots {\bf\boldsymbol{y}}^T_{2,M} \dots {\bf\boldsymbol{y}}^T_{M-1,M} \right]^T,
$$
conditioned on $\bf \boldsymbol{\theta}$, follows a  multivariate Gaussian distribution, i.e., $p({\bf Y''}  \vert \boldsymbol{\theta}) \sim \mathcal{CN}(\boldsymbol{\mu},\boldsymbol{\Sigma})$, with mean $\boldsymbol{\mu} =  \left[ \boldsymbol{\mu}^T_{1,2} \dots \boldsymbol{\mu}^T_{1,M} \boldsymbol{\mu}^T_{2,3} \dots \boldsymbol{\mu}^T_{2,M} \dots \boldsymbol{\mu}^T_{M-1,M} \right]^T$ and block diagonal covariance
\begin{equation}
\boldsymbol{\Sigma}
= {\rm diag}\left\{ \boldsymbol{\Sigma}_{1,2}, \cdots , \boldsymbol{\Sigma}_{1,M} , \boldsymbol{\Sigma}_{2,3}, \cdots , \boldsymbol{\Sigma}_{2,M} , \cdots , \boldsymbol{\Sigma}_{M-1,M}  \right\}.
\end{equation}
With that, we have
\begin{equation}
\label{eq:CRLBgen}
\left[  \textbf{I}( \boldsymbol{\theta} ) \right]_{i,j} =  \; \Tr \left\{  \boldsymbol{\Sigma}^{-1} \frac{ \partial \boldsymbol{\Sigma} }{ \partial \theta_i }  \boldsymbol{\Sigma}^{-1} \frac{ \partial \boldsymbol{\Sigma} }{ \partial \theta_j } \right\}  + 2 \operatorname{Re} \left\{  \frac{ \partial \boldsymbol{\mu}^H }{\partial \theta_i} \boldsymbol{\Sigma}^{-1}   \frac{ \partial \boldsymbol{\mu} }{\partial \theta_j}  \right\},
\end{equation}
with $1 \leq i \leq (4M-4)$ and $1 \leq j \leq (4M-4)$. The remaining computations of $\left[ \textbf{I}( \boldsymbol{\theta} ) \right]_{i,j}$ are straightforward and thus omitted. We note that without the convention of $t_\textit{ref}=r_\textit{ref}=1$ - and thus  $\boldsymbol{\theta}$ is a $4M \times 1$ vector instead - it can be shown that the map $\boldsymbol{\theta}  \mapsto \boldsymbol{\mu}$  is not injective which renders $\textbf{I}( \boldsymbol{\theta} )$ not invertible. Thus, the convention of reference antenna is needed to be able to compute the CRLB.







\section*{Appendix C - Closed-form Unpenalized ML estimator for Linear Arrays }
Here we derive the closed-form unpenalized (i.e. $\epsilon=0$) ML estimator for the linear array setup  described in Sec. \ref{sec:LinearArray}.  By leaving out the terms that do not depend on $\bf c$, it follows that, after a few manipulations, the optimization problem of (\ref{eq:EqModel5}) can be written as 
\iftoggle{2Collumn}{%
\begin{align}
\label{eq:HardCostLinear}
\left\{ \hat{c}_m \right\}  = & \arg \max_{ {\bf c} }   { {\bf Y'}^H   \bf C_{\rm eq}({\bf c })}  {\bf C}_{\rm eq}^{ \dagger}( {\bf c}){\bf Y'}  \nonumber \\ 
= & \arg \max_{ \left\{ c_m \right\} }  \sum_{\ell=1}^{M-1‎} f_{\rm L}(c_\ell,c_{\ell+1},{\bf \boldsymbol{y}}_{\ell+1,\ell}),
\end{align}
with
$$f_{\rm L}(c_\ell,c_{\ell+1},{\bf \boldsymbol{y}}_{\ell+1,\ell}) = {\bf \boldsymbol{y}}_{\ell+1,\ell}^{ H} {\bf \boldsymbol{\bar{c}}}_{\ell,\ell+1} {\bf \boldsymbol{\bar{c}}}_{\ell,\ell+1}^H {\bf \boldsymbol{y}}_{\ell+1,\ell} / {\bf \boldsymbol{\bar{c}}}_{\ell,\ell+1}^H {\bf \boldsymbol{\bar{c}}}_{\ell,\ell+1}.$$ See  (\ref{eq:smallC}) for structure of ${\boldsymbol{\bar{c}}}_{\ell,\ell+1}$, and (\ref{eq:subCov}) for structure of ${\bf \boldsymbol{y}}_{m,n}$. 
}{%
\begin{equation}
\label{eq:HardCostLinear}
\left\{ \hat{c}_m \right\}  = \arg \max_{ {\bf c} }   { {\bf Y'}^H   \bf C_{\rm eq}({\bf c })}  {\bf C}_{\rm eq}^{ \dagger}( {\bf c}){\bf Y'}    = \arg \max_{ \left\{ c_m \right\} }  \sum_{\ell=1}^{M-1‎} f_{\rm L}(c_\ell,c_{\ell+1},{\bf \boldsymbol{y}}_{\ell+1,\ell}),
\end{equation} 
with
$f_{\rm L}(c_\ell,c_{\ell+1},{\bf \boldsymbol{y}}_{\ell+1,\ell}) = {\bf \boldsymbol{y}}_{\ell+1,\ell}^{ H} {\bf \boldsymbol{\bar{c}}}_{\ell,\ell+1} {\bf \boldsymbol{\bar{c}}}_{\ell,\ell+1}^H {\bf \boldsymbol{y}}_{\ell+1,\ell} / {\bf \boldsymbol{\bar{c}}}_{\ell,\ell+1}^H {\bf \boldsymbol{\bar{c}}}_{\ell,\ell+1}.$ See  (\ref{eq:smallC}) for structure of ${\boldsymbol{\bar{c}}}_{\ell,\ell+1}$, and (\ref{eq:subMean}) for structure of ${\bf \boldsymbol{y}}_{m,n}$. 
}
%
Our ability to solve (\ref{eq:HardCostLinear}) is due to the following property.

\textit{Property 1:} For the function $f_{\rm L}(c_\ell,c_{\ell+1},{\bf\boldsymbol{y}}_{\ell+1,\ell})$, the maximum over $c_{\ell+1}$ equals $||{\bf\boldsymbol{y}}_{\ell+1,\ell}||^2$, and thus it does not depend on $c_\ell$.

Hence, the ML estimate of $c_{\ell+1}$, i.e. $\hat{c}_{\ell+1}$, can be found for a given  $c_{\ell}$. With that, the joint maximization problem (\ref{eq:HardCostLinear}) can be split into
\begin{equation*}
\label{eq:HardCostLinear3}
\hat{c}_{\ell+1} = \arg \max_{ x } f_{\rm L}(\hat{c}_\ell,x,{\bf\boldsymbol{y}}_{\ell+1,\ell}).
\end{equation*}
This optimization is a particular case of the Rayleigh quotient problem, and the solution is given in (\ref{eq:closedFormML}) when the reference element (i.e., the starting point) is chosen to be $c_1$.

We now provide a short proof for Corollary 1. For the case of linear arrays with coupling solely between adjacent antennas, the optimization problem in (\ref{eq:GMMeCost}) can be written - ignoring any constraint for now - as
\begin{equation}
\label{eq:HardCostLineard}
\hat{\boldsymbol{c}}_{\rm GMM}=  \arg \min_{ \substack{\boldsymbol{c}  }}  \sum_{\ell=1}^{M-1‎} f_{\rm G}(c_\ell,c_{\ell+1},{\bf \boldsymbol{y}}_{\ell+1,\ell})
\end{equation}
where $ f_{\rm G}(c_\ell,c_{\ell+1},{\bf \boldsymbol{y}}_{\ell+1,\ell})=|y_{\ell+1,\ell} c_{\ell+1} -y_{\ell,\ell+1} c_{\ell}|^2$. We solve (\ref{eq:HardCostLineard}) using the following property.

\textit{Property 2:} Letting $\hat{c}_{\ell}$ be the ML estimator from (\ref{eq:closedFormML}), it follows that
\begin{equation}
\label{eq:HardCostLineardd}
f_{\rm G}(\hat{c}_{\ell} ,\hat{c}_{\ell+1},{\bf \boldsymbol{y}}_{\ell+1,\ell} )=0, \; \forall \ell.
\end{equation}
Thus, the GMM  solution (under any of the $2$ constraints)  coincides with that of the ML  up to a common complex scalar. Uniqueness follows since the GMM cost function is quadratic.





\begin{thebibliography}{10}
\providecommand{\url}[1]{#1}
\csname url@samestyle\endcsname
\providecommand{\newblock}{\relax}
\providecommand{\bibinfo}[2]{#2}
\providecommand{\BIBentrySTDinterwordspacing}{\spaceskip=0pt\relax}
\providecommand{\BIBentryALTinterwordstretchfactor}{4}
\providecommand{\BIBentryALTinterwordspacing}{\spaceskip=\fontdimen2\font plus
\BIBentryALTinterwordstretchfactor\fontdimen3\font minus
  \fontdimen4\font\relax}
\providecommand{\BIBforeignlanguage}[2]{{%
\expandafter\ifx\csname l@#1\endcsname\relax
\typeout{** WARNING: IEEEtran.bst: No hyphenation pattern has been}%
\typeout{** loaded for the language `#1'. Using the pattern for}%
\typeout{** the default language instead.}%
\else
\language=\csname l@#1\endcsname
\fi
#2}}
\providecommand{\BIBdecl}{\relax}
\BIBdecl

\bibitem{LUP2174140}
F.~Rusek, D.~Persson, B.~K. Lau, E.~G. Larsson, T.~L. Marzetta, O.~Edfors, and
  F.~Tufvesson, ``{Scaling up MIMO: opportunities and challenges with very
  large arrays},'' \emph{IEEE Signal Processing Magazine}, vol.~30, no.~1, pp.
  40--60, 2013.

\bibitem{LUP4305564}
E.~G. Larsson, O.~Edfors, F.~Tufvesson, and T.~L. Marzetta, ``{Massive MIMO for
  next generation wireless systems},'' \emph{IEEE Communications Magazine},
  vol.~52, no.~2, pp. 186--195, 2014.

\bibitem{978730}
L.~Zheng and D.~N.~C. Tse, ``{Communication on the Grassmann manifold: a
  geometric approach to the noncoherent multiple-antenna channel},'' \emph{IEEE
  Transactions on Information Theory}, vol.~48, no.~2, pp. 359--383, Feb 2002.

\bibitem{Marzetta}
T.~L. Marzetta, ``Noncooperative cellular wireless with unlimited numbers of
  base station antennas,'' \emph{IEEE Transactions on Wireless Communications},
  vol.~9, no.~11, pp. 3590--3600, Nov. 2010.

\bibitem{Proakis}
M.~Salehi and J.~Proakis, \emph{Digital Communications}.\hskip 1em plus 0.5em
  minus 0.4em\relax McGraw-Hill Education, 2007.

\bibitem{Balanis}
C.~A. Balanis, \emph{Antenna Theory: Analysis and Design}.\hskip 1em plus 0.5em
  minus 0.4em\relax Wiley-Interscience, 2005.

\bibitem{5722383}
F.~Kaltenberger, H.~Jiang, M.~Guillaud, and R.~Knopp, ``{Relative channel
  reciprocity calibration in MIMO/TDD systems},'' in \emph{2010 Future Network
  and Mobile Summit}, June 2010.

\bibitem{6573241}
{M. Petermann et al.}, ``{Multi-User Pre-Processing in Multi-Antenna OFDM TDD
  Systems with Non-Reciprocal Transceivers},'' \emph{IEEE Transactions on
  Communications}, vol.~61, no.~9, pp. 3781--3793, September 2013.

\bibitem{966592}
K.~Nishimori, K.~Cho, Y.~Takatori, and T.~Hori, ``{Automatic calibration method
  using transmitting signals of an adaptive array for TDD systems},''
  \emph{IEEE Transactions on Vehicular Technology}, vol.~50, no.~6, pp.
  1636--1640, Nov 2001.

\bibitem{6963664}
K.~Nishimori, T.~Hiraguri, T.~Ogawa, and H.~Yamada, ``{Effectiveness of
  implicit beamforming using calibration technique in massive MIMO system},''
  in \emph{2014 IEEE International Workshop on Electromagnetics (iWEM)}, Aug
  2014, pp. 117--118.

\bibitem{7416977}
X.~Luo, ``{Robust Large Scale Calibration for Massive MIMO},'' in \emph{2015
  IEEE Global Communications Conference (GLOBECOM)}, Dec 2015.

\bibitem{7239634}
H.~Wei, D.~Wang, H.~Zhu, J.~Wang, S.~Sun, and X.~You, ``{Mutual Coupling
  Calibration for Multiuser Massive MIMO Systems},'' \emph{IEEE Transactions on
  Wireless Communications}, vol.~15, no.~1, pp. 606--619, Jan 2016.

\bibitem{6760595}
{R. Rogalin et al.}, ``{Scalable Synchronization and Reciprocity Calibration
  for Distributed Multiuser MIMO},'' \emph{IEEE Transactions on Wireless
  Communications}, vol.~13, no.~4, April 2014.

\bibitem{ArgosThesis}
{C. Shepard et al.}, ``Argos: Practical many-antenna base stations,'' in
  \emph{Proceedings of the 18th Annual International Conference on Mobile
  Computing and Networking}, ser. Mobicom '12.\hskip 1em plus 0.5em minus
  0.4em\relax New York, NY, USA: ACM, 2012, pp. 53--64.

\bibitem{6502966}
R.~Rogalin, O.~Y. Bursalioglu, H.~C. Papadopoulos, G.~Caire, and A.~F. Molisch,
  ``Hardware-impairment compensation for enabling distributed large-scale
  mimo,'' in \emph{{Information Theory and Applications Workshop (ITA), 2013}},
  Feb 2013.

\bibitem{Wei2015}
H.~Wei, D.~Wang, J.~Wang, and X.~You, ``{TDD reciprocity calibration for
  multi-user massive MIMO systems with iterative coordinate descent},''
  \emph{Science China Information Sciences}, vol.~59, no.~10, p. 102306, 2015.

\bibitem{7032189}
H.~Papadopoulos, O.~Bursalioglu, and G.~Caire, ``{Avalanche: Fast RF
  calibration of massive arrays},'' in \emph{2014 IEEE Global Conference on
  Signal and Information Processing (GlobalSIP)}, Dec 2014, pp. 607--611.

\bibitem{J7037384}
J.~Vieira, F.~Rusek, and F.~Tufvesson, ``{Reciprocity calibration methods for
  massive MIMO based on antenna coupling},'' in \emph{2014 IEEE Global
  Communications Conference (GLOBECOM)}, Dec 2014, pp. 3708--3712.

\bibitem{7340979}
H.~Wei, D.~Wang, and X.~You, ``{Reciprocity of mutual coupling for TDD massive
  MIMO systems},'' in \emph{2015 International Conference on Wireless
  Communications Signal Processing (WCSP)}, Oct 2015.

\bibitem{EURECOM+4463}
{J. Xiwen et al.}, ``{MIMO}-{TDD} reciprocity under hardware imbalances:
  {E}xperimental results,'' in \emph{2015 {IEEE} {I}nternational {C}onference
  on {C}ommunications {ICC}, 8-12 {J}une 2015, {L}ondon, {U}nited {K}ingdom},
  {L}ondon, {U.K.}, 2015.

\bibitem{1142529}
R.~Jedlicka, M.~Poe, and K.~Carver, ``Measured mutual coupling between
  microstrip antennas,'' \emph{IEEE Transactions on Antennas and Propagation},
  vol.~29, no.~1, pp. 147--149, Jan 1981.

\bibitem{BasestationPaper}
{J. Vieira et al.}, ``{A flexible 100-antenna testbed for Massive MIMO},'' in
  \emph{IEEE GLOBECOM 2014 Workshop on Massive MIMO: from theory to practice},
  Dec 2014.

\bibitem{hall2004generalized}
A.~Hall, \emph{Generalized Method of Moments}, ser. Advanced Texts in
  Econometrics.\hskip 1em plus 0.5em minus 0.4em\relax OUP Oxford, 2004.

\bibitem{Bishop}
C.~M. Bishop, \emph{Pattern Recognition and Machine Learning (Information
  Science and Statistics)}.\hskip 1em plus 0.5em minus 0.4em\relax Secaucus,
  NJ, USA: Springer-Verlag New York, Inc., 2006.

\bibitem{Hoerl}
A.~E. Hoerl and R.~W. Kennard, ``Ridge regression: Biased estimation for
  nonorthogonal problems,'' \emph{Technometrics}, vol.~42, no.~1, pp. 80--86,
  Feb. 2000.

\bibitem{scharf1991statistical}
L.~Scharf and C.~Demeure, \emph{Statistical Signal Processing: Detection,
  Estimation, and Time Series Analysis}, ser. Addison-Wesley series in
  electrical and computer engineering.\hskip 1em plus 0.5em minus 0.4em\relax
  Addison-Wesley Publishing Company, 1991.

\bibitem{Kay}
S.~M. Kay, \emph{Fundamentals of Statistical Signal Processing: Estimation
  Theory}.\hskip 1em plus 0.5em minus 0.4em\relax Prentice-Hall, Inc., 1993.

\bibitem{Haykin}
S.~Haykin, \emph{Adaptive Filter Theory (3rd Ed.)}.\hskip 1em plus 0.5em minus
  0.4em\relax Upper Saddle River, NJ, USA: Prentice-Hall, Inc., 1996.

\bibitem{molisch2010wireless}
A.~Molisch, \emph{Wireless Communications}, ser. Wiley - IEEE.\hskip 1em plus
  0.5em minus 0.4em\relax Wiley, 2010.

\bibitem{Paulraj}
A.~Paulraj, R.~Nabar, and D.~Gore, \emph{Introduction to Space-Time Wireless
  Communications}, 1st~ed.\hskip 1em plus 0.5em minus 0.4em\relax New York, NY,
  USA: Cambridge University Press, 2008.

\bibitem{7434053}
X.~Luo, ``{Multi-User Massive MIMO Performance with Calibration Errors},''
  \emph{IEEE Transactions on Wireless Communications}, vol.~15, no.~7, pp.
  4521--4534, Jul 2016.

\bibitem{7018038}
{W. Zhang et al.}, ``{Large-Scale Antenna Systems With UL/DL Hardware Mismatch:
  Achievable Rates Analysis and Calibration},'' \emph{IEEE Transactions on
  Communications}, vol.~63, no.~4, pp. 1216--1229, April 2015.

\bibitem{schenk2008rf}
T.~Schenk, \emph{RF Imperfections in High-rate Wireless Systems: Impact and
  Digital Compensation}.\hskip 1em plus 0.5em minus 0.4em\relax Springer, 2008.

\bibitem{van2004detection}
H.~Van~Trees, \emph{Detection, Estimation, and Modulation Theory}.\hskip 1em
  plus 0.5em minus 0.4em\relax Wiley, 2004.

\bibitem{daniel1990applied}
W.~Daniel, \emph{Applied nonparametric statistics}, ser. The Duxbury advanced
  series in statistics and decision sciences.\hskip 1em plus 0.5em minus
  0.4em\relax PWS-Kent Publ., 1990.

\end{thebibliography}


\end{document}